\documentclass{emulateapj}

\begin{document}

\title{The mass-metallicity and luminosity-metallicity relations from DEEP2 at $\MakeLowercase{z}\sim0.8$}
\author{H. J. Zahid, L. J. Kewley, F. Bresolin}
\affil{Institute for Astronomy 2680 Woodlawn Dr., Honolulu, HI 96822}

\date{January 28, 2011}
\submitted{Accepted Version}
                                      % Activate to display a given date or no date
%\begin{document}
\begin{abstract}

We present the mass-metallicity (MZ) and luminosity-metallicity (LZ) relations at $z\sim0.8$ from 
$\sim 1350$ galaxies in the Deep Extragalactic Evolutionary Probe 2 (DEEP2) survey.  We determine stellar masses by fitting the spectral energy distribution inferred from photometry with current stellar population synthesis models.  This work raises the number of galaxies with metallicities at z$\sim 0.8$ by more than an order of magnitude. We investigate the evolution in the MZ and LZ relations in comparison with local MZ and LZ relations determined in a consistent manner using $\sim 21,000$ galaxies in the Sloan Digital Sky Survey.    We show that high stellar mass galaxies ($M \sim 10^{10.6} M_\odot$) at z$\sim 0.8$ have attained the chemical enrichment seen in the local universe, while lower stellar mass galaxies ($M \sim 10^{9.2} M_\odot$) at z$\sim 0.8$ have lower metallicities ($\Delta {\rm log(O/H)} \sim 0.15$~dex) than galaxies at the same stellar mass  in the local universe.  We find that the LZ relation evolves in both metallicity and B-band luminosity between z$\sim 0.8$ and $z\sim 0$, with the B-band luminosity evolving as a function of stellar mass.  We emphasize that the B-band luminosity should not be used as a proxy for stellar mass in chemical evolution studies of star-forming galaxies.  Our study shows that both the metallicity evolution and the B-band luminosity evolution for emission-line galaxies between the epochs are a function of stellar mass, consistent with the cosmic downsizing scenario of galaxy evolution.

\end{abstract}

%\section{abstract}

\section{Introduction}

The heavy-element abundance in galaxies is a key physical property for understanding galaxy evolution. Metals are formed in massive stars and are dispersed into the interstellar medium (ISM) via mass-loss processes. Hence, metallicity provides an important record of the star formation history of a galaxy. Galaxies and their chemical abundances, however, do not evolve as closed systems where metals are dispersed into the ISM as gas is converted into stars. Rather, they are modulated by inflow of unenriched gas \citep[e.g.,][]{Kewley2010} and a complex "feedback" mechanism where supernovae and stellar winds influence the ISM by inducing outflows of gas into the intergalactic medium (IGM) and regulating star formation through reheating \citep{Larson1974, Larson1975, Kauffmann1998, Somerville1999, Springel2003}. 

A simple "closed-box" model in which feedback is not considered predicts a correlation between two fundamental parameters$-$mass and metallicity \citep{vandenbergh1962, Schmidt1963, Searle1972, Erb2006}. The pioneering work of \citet{Lequeux1979} first showed this correlation in local irregular galaxies. Many subsequent efforts have focused on the relation between luminosity and metallicity \citep{Skillman1989, Zaritsky1994, Garnett1997, Lamareille2004, Salzer2005} where luminosity is taken as a proxy for mass because of the difficulty in inferring galaxy masses from observables. Early attempts to extend this relation to intermediate redshifts found that the luminosity-metallicity (LZ) relation at earlier epochs was consistent with the local universe \citep{Kobulnicky1999, Carollo2001}. However, many recent investigations have provided compelling evidence for evolution of the LZ relation over cosmological times \citep{Kobulnicky2000, Pettini2001, Kobulnicky2003a, Kobulnicky2004, Maier2004, Shapley2004, Shapley2005}. Luminosities of galaxies evolve along with their chemical abundances on cosmological timescales and disentangling the contribution for each has posed difficulties in these studies.

Evolution in the LZ relation implies an evolution in the more fundamental mass-metallicity (MZ) relation. Development of more sophisticated models for stellar population synthesis \citep{Bruzual2003} and gaseous nebula \citep{Ferland1996, Charlot2001} allowed \citet[hereafter T04]{Tremonti2004} to establish and quantify the local MZ relation for  $\sim\!\!53,000$ star-forming galaxies from the Sloan Digital Sky Survey (SDSS). They found that metallicity increases linearly with stellar mass for galaxies having masses between $10^{8.5} M_{\odot}$ and $10^{10.5} M_{\odot}$  and flattens out at higher masses. They attribute this depletion of metals in less massive galaxies to ubiquitous galactic winds that strip metals more effectively from galaxies with shallow potential wells, dispersing them into the IGM. Alternatively, it has been suggested that low mass systems have ongoing star formation and have yet to convert much of their gas into stars and therefore are less evolved, less metal-rich systems as compared to more massive galaxies which have undergone rapid star formation \citep{Brooks2007, DeRossi2007, Mouchine2008}. Finally, a varying galaxy-integrated initial mass function has also emerged as a possible explanation for the observed MZ relation \citep{Koppen2007}.

Cosmological hydrodynamic simulations \citep{Brooks2007, Oppenheimer2008} have predicted an evolution in the MZ relation and recent observations at intermediate redshifts ($z\sim0.7$) have provided strong evidence for an evolution. Using the Gemini Deep Deep Survey, \citet{Savaglio2005} first showed the MZ relation beyond the local universe for galaxies at $z = 0.4-1.0$.  Their relation intersects the local relation at $M \sim 10^{10.4}M_{\odot}$ but is $\sim0.3$ dex lower at $M \sim 10^{9.3}M_{\odot}$, implying a much steeper slope.  They conclude that the slope of the MZ relation must become flatter over time. \citet{Cowie2008} found a difference $>0.2$ dex for a redshift range of $0.475<z<0.9$ with a similar slope. More recently, \citet{Lamareille2009} find a flatter slope in the MZ relation out to $z \! \sim \! 0.9$ and an average difference in metallicity of $\sim\!0.2$ dex. To date, the MZ relation has been extended beyond $z\!\sim\!3$ \citep{Shapley2005, Erb2006, Maiolino2008,Perez-Montero2009,Mannucci2009}. All these studies have shown evidence of evolution but rigorous quantitative analysis has been hampered by disparate results and small sample sizes leaving the MZ relation at higher redshifts quantitatively uncertain.

In this paper we present results of our study determining the mass-metallicity and luminosity-metallicity relation for galaxies in the redshift range of $0.75<z<0.82$. We infer metallicities from gas phase oxygen abundances using strong line diagnostics \citep[hereafter KE08]{Pagel1979, Kewley2002, Kewley2008}. We make use of publicly available data from the Deep Extragalactic Evolutionary Probe 2 \citep[DEEP2,][]{Davis2003}. We measure the MZ and LZ relations from 1,348 galaxies, a substantial increase in the number of objects studied in this redshift range to date. We note that a large sample of galaxies with metallicity and stellar mass determinations in such small redshift range is ideal for reliably constraining the MZ relation at a specific redshift. In $\S \, 2$ we present our criteria for selection of galaxies and properties of our selected sample. In $\S \, 3$ we describe our method for determining masses, our correction for underlying Balmer absorption and our method for determining masses. In $\S \, 4$ we present our derived MZ  and LZ relations.  In $\S \, 5$ we present a discussion of our results. Here we compare the MZ and LZ relation and infer the B-band luminosity and evolution as a function of stellar mass and we compare our MZ relation with relations derived by other authors at similar redshfits. We summarize our results in $\S\, 6$. Where necessary, we adopt the standard cosmology $(H_{0}, \Omega_{m}, \Omega_{\Lambda}) = (70\, km\, s^{-1}\, Mpc^{-1}, 0.3, 0.7)$.

\section{The Data}

%In this section we discuss the DEEP2 data ($\S \, 2.1$), our selected sample ($\S \, 2.2$) and its properties ($\S \, 2.3$).  - unnecessary

\subsection{DEEP2 Survey Data}

Our $z\sim0.8$ sample comes from the DEEP2 (Deep Extragalactic Evolutionary Probe 2) survey. The DEEP2 team used the DEIMOS \citep{Faber2003} multi-object spectrograph on the Keck telescope to measure spectra for galaxies in 4 fields covering 3.5 $deg^{2}$ down to a limiting magnitude of $R_{AB} = 24.1$. The survey contains spectra for $\sim\!\!50,000$ galaxies, the majority of which are in the redshift range $0.7<z<1.4$. The relatively high-resolution ($R\ge5000$) spectra cover the nominal spectral range 6500-9100$\AA$. For this study, we used the third data release\footnote{http://www.deep.berkeley.edu/DR3/} data and the 1-d spectra were obtained from the Horne extraction \citep{Horne1986}.

BRI photometry was obtained by the DEEP2 team for these objects from imaging carried out using the CFH12K camera on the 3.6 m Canada France Hawaii Telescope \citep{Coil2004}. For a subsample of the data, $K_s$-band photometry was available in addition to the BRI photometry. These data were obtained using the Wide Field Infrared Camera on the 5m Hale telescope at Mt. Palomar \citep{Bundy2006}. All magnitudes in this study are in the AB system.

\subsection{Data Selection}

\begin{deluxetable*}{lrrr}

\tablewidth{475pt}

\tablecaption{Sample Selection}

\tablehead{\colhead{Selection Criteria} & \colhead{Total} &\colhead{Blue} & \colhead{Red} } 

\startdata
Total number of unique objects    & 46,337 & - & -\\
Galaxies with secure redshift determination & 31,656 & - & -\\
Galaxies with redshifts $0.75<z<0.82$ & 4,198 & 3,323 & 875 \\
Rest-frame wavelength cover $3720-5020\AA$ & 2,738 & 2,152 & 586\\
Measured metallicities & 2,504 & 2,117 & 387\\
SN $H\beta > 3$ & 2,168 & 1,996 & 172 \\
$EW(H\beta) > 4\AA$ & 1,973 & 1,902 & 71 \\
$\sigma_{R23} < 2$ & 1,751 & 1,702 & 49 \\
Robust continuum & 1,713 & 1,665 & 48\\
Not AGN & 1,679 & 1,631 & 48\\
Blue star forming galaxies   & 1,631 & 1,631 & 0\\
\enddata

\label{tab:cut}

\tablecomments{The first column gives the particular selection criteria and the second column gives the total number of galaxies that meet this and all criteria above it in the table. The third and fourth column splits the total number of galaxies into blue and red sub-samples as defined by equation \ref{eq:color}.}

\end{deluxetable*}

\begin{figure*}
\includegraphics[width=7 in]{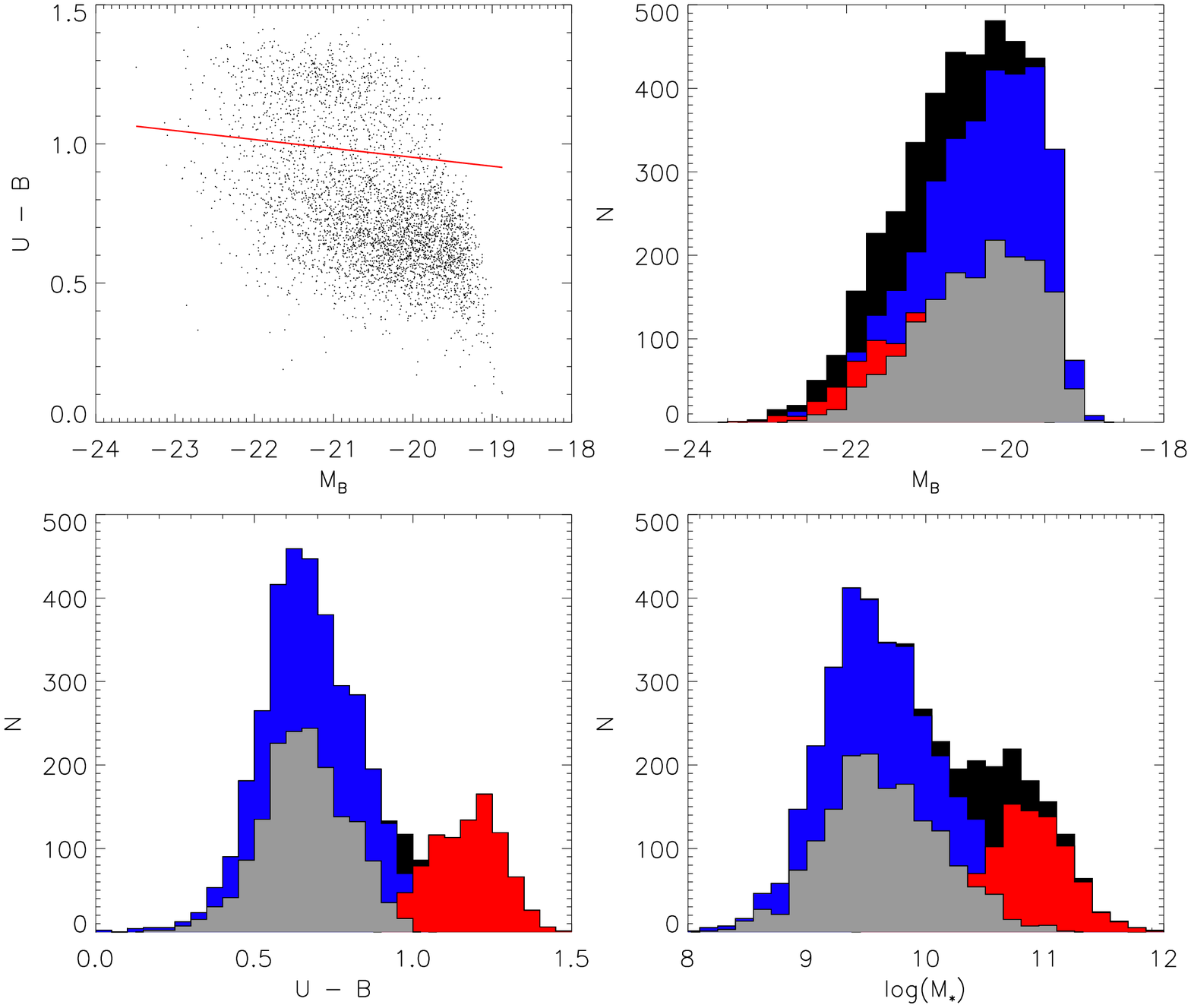}
\caption{The top left panel shows the color magnitude diagram of the full sample of 4,198 galaxies with the required wavelength coverage from which we make our selection. The red line is the blue-red color division given by equation \ref{eq:color}. The black histogram in each of the other three panels is for the full sample. The blue and red histograms are for the blue and red galaxies within the full sample and the grey histogram is for our selected sub-sample of 1,631 galaxies. The top right panel shows the distribution of the absolute B-band magnitude. The bottom left panel shows the color distribution and bottom right panel shows the mass distribution.}
\label{fig:br_hist}
\end{figure*}

From the 46,337 unique objects in the survey, we begin by selecting 31,656 galaxies for which we have secure redshift determinations from the DEEP2 team ($Q\geq 3$, see website for more details). For our determination of metallicity, we use the $R_{23}$ strong line diagnostic, first calibrated by \citet{Pagel1979}. We apply a more recent $R_{23}$ calibration based on the stellar population synthesis and photoionization model calibrations from \citet{Kewley2002}, and updated in \citet{Kobulnicky2004}. In the text, we refer to this calibration as KK04. The $R_{23}$ method is a parameterization of metallicity as function of the ratio of the oxygen nebular emission ([OII]$\lambda 3727$ doublet, [OIII]$\lambda 4959$ and [OIII]$\lambda5007$) to $H\beta$. The method is formally defined in $\S \, 3.3$.  Given the nominal spectral range of the DEEP2 survey, we can sufficiently bracket the required emission lines within the $0.75<z<0.82$ redshift range.  Due to a varying object position on the slitmask, the actual spectral range varies from object to object and we require that the spectra cover the wavelength range of $3720-5020\AA$, leaving us with  2,738 galaxies. We were unable to measure the metallicity in 234 galaxies due to our inability to fit the oxygen or $H\beta$ emission lines, leaving us with 2,504 galaxies.

We calculate the  $R_{23}$ ratios using the continuum normalized flux (see Appendix for full description).  In the appendix we show that the pseudo-EW is equal, to within the errors, to the equivalent width and refer to it hereafter as the EW.  The errors on our equivalent width and line ratio determination come from propagating the measurement uncertainties in the spectrum.  In selecting galaxies for analysis, we require that the S/N $H{\beta}>3$, $EW(H\beta) > 4\AA$ and the error in the $R_{23}$ line ratio be less than 2. At high metallicities, the line strength of the [OII] and [OIII] line diminishes due to the effectiveness of line cooling of the HII region. In order to avoid biasing our MZ relation to lower metallicities we do not place a S/N cut on these lines. We normalize the line flux to the continuum by fitting a polynomial to the underlying continuum and dividing the spectrum by this fit. We require that the fit to the continuum be $1\sigma_{c}$ above zero, where $\sigma_{c}$ is the standard deviation of the line-subtracted spectrum (see appendix for more detail). These cuts leave us with 1,713 galaxies.

Strong-line methods of metallicity determination assume a star forming population of galaxies and are not applicable to AGN. We identify all galaxies with $log(R_{23}) > 1$ as AGN and remove them from our sample (Kewley et al., in prep). Furthermore, using the color separation for blue and red galaxies parameterized by \citet{Willmer2006} for the DEEP2 survey, \citet{Weiner2007} conclude that a very small fraction of blue galaxies have AGN whereas at least half of the red galaxies with detectable emission lines show evidence of AGN emission. Similar to T04, we select the blue star forming galaxies to analyze, as described in the following section. We remove the red galaxies from our sample and are left with 1,631 blue emission line galaxies. The first and second column of Table \ref{tab:cut} summarizes our selection. The median for $EW(H\beta)$, S/N $H\beta$ and $\sigma_{R23}$ of our selected sample is 12.1, 10.0 and 0.65 respectively.

\subsection{Properties of Selected Sample}

\begin{deluxetable}{lccc}

%\tablewidth{475pt}

%% Keep a portrait orientation
\tablewidth{\columnwidth}

%% Over-ride the default font size
%% Use Default (12pt)

%% Use \tablewidth{?pt} to over-ride the default table width.
%% If you are unhappy with the default look at the end of the
%% *.log file to see what the default was set at before adjusting
%% this value.

%% This is the title of the table.
\tablecaption{Median Properties of the DEEP2 Galaxies}
%% This command over-rides LaTeX's natural table count
%% and replaces it with this number.  LaTeX will increment 
%% all other tables after this table based on this number
%\tablenum{1}

%% The \tablehead gives provides the column headers.  It
%% is currently set up so that the column labels are on the
%% top line and the units surrounded by ()s are in the 
%% bottom line.  You may add more header information by writing
%% another line between these lines. For each column that requries
%% extra information be sure to include a \colhead{text} command
%% and remember to end any extra lines with \\ and include the 
%% correct number of &s.\
 
%\tablewidth{10 cm}

\tablehead{{Sample} & \colhead{$U-B$} &\colhead{$M_{B}$} & \colhead{Mass} \\
&&&\colhead{ $log(M/M_{\odot})$} } 

%% All data must appear between the \startdata and \enddata commands
\startdata
DEEP2 Full            & 0.72 & -20.4 & 9.80 \\
DEEP2 Red           & 1.18 & -21.1 & 10.85 \\
DEEP2 Blue          & 0.67 & -20.2 & 9.62 \\
DEEP2 Selected   & 0.66 & -20.3 & 9.59 \\
%Selected SDSS     & 0.80 & -19.6 & 9.84 \\
\enddata

%% Include any \tablenotetext{key}{text}, \tablerefs{ref list},
%% or \tablecomments{text} between the \enddata and 
%% \end{deluxetable} commands
\label{tab:median}
%% General table comment marker

%\tablewidth{\columnwidth}
\tablecomments{The median properties of the data. In the first column we identify the four samples. The next three columns give the median $U-B$ color, absolute B-band magnitude and mass for each of these samples.}
%\tablecomments{RMS deviation comparison of one-to-one agreement from \citet{Kobulnicky2003b} (top row), our automated method (middle row) and the mean of the errors determined from the statistical measurement uncertainties of the spectra (bottom row).}
\end{deluxetable}

\citet{Willmer2006} studied the red and blue galaxy luminosity functions by dividing the DEEP2 sample at the trough of the bimodal $U-B$ color distribution. The U- and B-band absolute magnitudes are synthesized by relating the observed magnitudes to spectral energy distributions (SEDs) of nearby galaxies and inferring a rest-frame absolute magnitude from this SED. We follow a similar procedure and synthesize the U- and B-band absolute magnitude using the filters of \citet{Buser1978} and \citet{Azusienis1969} respectively. The red and blue galaxy color division found by \citet{Willmer2006} is given by
\begin{equation}
U-B = 1.0 - 0.032(M_{B} + 21.5).
\label{eq:color}
\end{equation}
The color division has been converted to AB magnitudes using Table 1 from \citet{Willmer2006}. In column three and four of Table \ref{tab:cut} we examine the cuts from our selection with respect to the galaxy color.

The top left panel of Figure \ref{fig:br_hist} shows the color-magnitude diagram and the solid red line is the color division given by equation \ref{eq:color}. The other three panels give the absolute B-band magnitude, stellar mass (derived in $\S \, 3.1$) and $U-B$ color distribution. The black histogram gives the full sample of 4,198 DEEP2 galaxies between $0.75<z<0.82$. We refer to this sample as the DEEP2 Full sample. Using the $U-B$ color division, we split the DEEP2 Full sample into sub-samples of red and blue galaxies.  The histograms of these sub-samples are plotted in red and blue and we refer to these sub-samples as the DEEP2 Red and DEEP2 Blue samples respectively. The grey histogram is the distribution of our selected sample of 1,631 galaxies described in $\S \, 2.2$ and we refer to this as our DEEP2 Selected sample. The median properties of these samples are given in Table \ref{tab:median}.

The absolute B-band magnitude distribution in the top right panel of Figure \ref{fig:br_hist} shows a range of $-19.0>M_B>-23$ for the DEEP2 Full sample.  \citet{Faber2007} showed that luminosity functions derived for blue galaxies from the DEEP2 data have a $M_B^\ast = -21.15$ in our selected redshift range. Our DEEP2 Selected sample consists of galaxies around the knee of the blue galaxy luminosity function \citep{Schechter1976} with $\sim\!15\%$ (245/1631) of our galaxies brighter than $M_B^\ast$. The bottom left panel of Figure \ref{fig:br_hist} shows that the $U-B$ color range is $0<U-B<1.5$ for the DEEP2 Full sample. The color division between red and blue is at $U-B\sim1$ and the bimodal distribution is clearly highlighted by the well separated peaks in the blue and red histograms. The bimodal color distribution separates red early-type galaxies from blue late-type star forming galaxies \citep{Strateva2001, Hogg2003, Baldry2004, Willmer2006}. The mass distribution shown in the bottom right panel of Figure \ref{fig:br_hist} reflects this bimodality, whereby the red galaxies in our sample are not only brighter, but more massive and separated from the lower mass blue galaxies.

By comparing the histograms and median properties in Figure~\ref{fig:br_hist}, it is clear that the DEEP2 Red sample of galaxies are brighter and more massive than the DEEP2 Blue sample. However, when comparing the histograms and median properties of our DEEP2 Selected sample, which is comprised solely of blue galaxies, with the DEEP2 Blue sample, there is no selection bias towards the brighter or more massive galaxies as one may expect given the selection criteria. We conclude from the comparison of the properties of the DEEP2 Selected and DEEP2 Blue sample that our selected galaxies are representative of the population of blue star forming of galaxies in the DEEP2 survey at $0.75<z<0.8$.

\section{Data Analysis}

%In this section we will discuss our method of determining mass ($\S \, 3.1$), metallicity ($\S \, 3.2$) and our correction for underlying stellar Balmer absorption ($\S \, 3.3$).  unnecessary

\subsection{Stellar Mass Determination}

\begin{figure*}
\includegraphics[width=7 in]{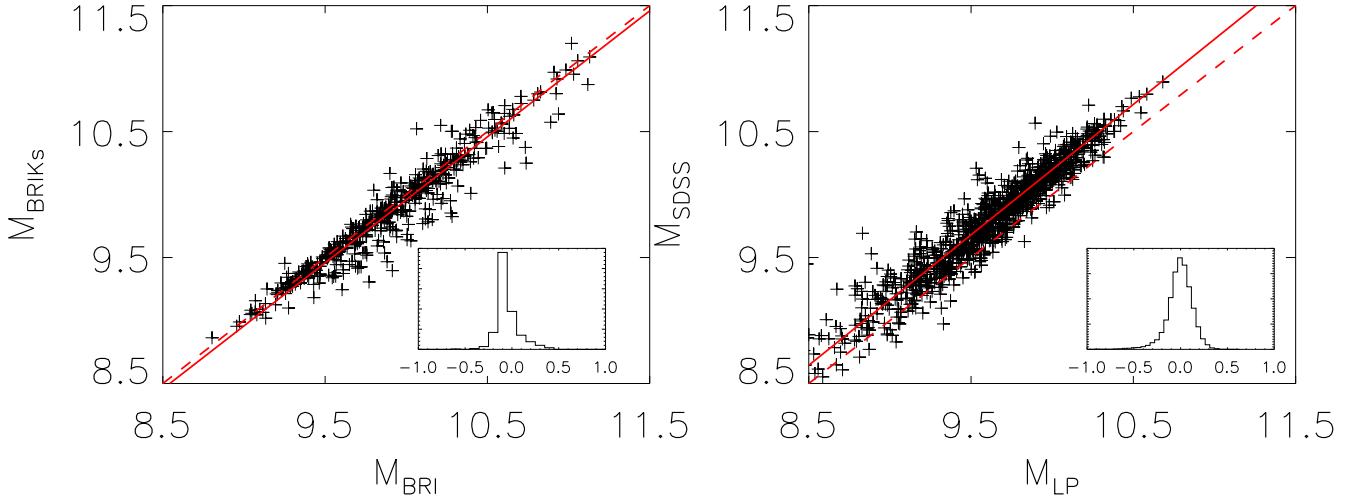}
\caption{Comparison of stellar masses estimates. The left panel shows the mass estimates using BRI-band photometry ($M_{BRI}$, x-axis) as compared with $BRIK_{s}$-band photometry ($M_{BRIKs}$, y-axis). The right panel compares mass estimates made using Le Phare ($M_{LP}$, x-axis) and those provided by the MPA/JHU team ($M_{SDSS}$, y-axis). The dashed line in each plot is the one-to-one correspondence of the two data sets. The solid line is a bisector fit to the data. The sub-panel in each of the plots shows a histogram of the differences between the two estimates with -0.04 and 0.19 dex added as corrections to the $M_{BRI}$ and $M_{LP}$ mass estimates respectively.}
\label{fig:mass}
\end{figure*}

We estimate galaxy stellar masses by comparing photometry with stellar population synthesis models in order to determine the mass-to-light ratio \citep{Bell2003, Fontana2004}. We use the Le Phare\footnote{$\url{http://www.cfht.hawaii.edu/{}_{\textrm{\symbol{126}}}arnouts/LEPHARE/cfht\_}$ $\url{lephare/lephare.html}$} code developed by Arnouts, S. \& Ilbert, O. to estimate the galactic stellar mass. Le Phare is a set of FORTRAN routines developed to determine the photometric redshifts of galaxies. If the redshift is known, it can be held fixed while other physical parameters of the galaxy are determined (i.e. stellar mass) by SED fitting.

The stellar templates of \citet{Bruzual2003} and an IMF described by \citet{Chabrier2003} are used to synthesize magnitudes. The 27 models span three metallicities and seven exponentially decreasing star formation models (SFR $\propto e^{-t/\tau}$) with $\tau = 0.1,0.3,1,2,3,5,10,15$ and $30$ Gyrs. We apply the extinction law of \citet{Calzetti2000} allowing E(B-V) to range from 0 to 0.6 and stellar population ages ranging from 0 to 13 Gyrs. \citet{Ilbert2009} provide a more detailed description of how the physical parameters are estimated. 

We are able to determine stellar masses using $BRI$ and $K_s$-band data for 454 of our selected galaxies. The K-band survey of the DEEP2 fields covers $60\%$ of the area so K-band photometry is not available for all galaxies in the survey \citep{Bundy2006}. In order to test for systematic variation in the stellar mass determination for the remaining sample of galaxies with only BRI photometry, we compare the stellar masses for our 454 galaxies determined with $BRI$ and $K_s$-band data with stellarmasses determined for the same galaxies with just the $BRI$ bands. The left panel of Figure \ref{fig:mass} shows the comparison. The x-axis is the mass determination from $BRI$ band data and the y-axis is the mass determination from $BRIK_s$-band data. The dashed line is the one-to-one correspondence and the solid line is a linear bisector fit to the data. 

We perform a bootstrap linear fit by taking the bisector of the X vs Y and Y vs X regression using the routine $\emph{boot\_xyfit.pro}$ from the IDL astronomy users library. Bootstrapping is a non-parametric statistical method of inferring errors whereby errors are calculated from the distribution of the fitted parameters for many fits to randomly selected subsample of the data. In order to minimize the covariance between the slope and intercept, we zero point the data by subtracting 10 from the logarithm of the stellar mass. The linear fit is given by
\begin{equation}
X_{BRIK_s} = (-0.042\pm0.006) - (1.00\pm0.01) X_{BRI} , 
\label{eq:ks}
\end{equation}
where $X_{BRI} = M_{BRI} - 10, X_{BRIK_s} = M_{BRIK_s} - 10 $ and $M_{BRI}$ and $M_{BRIK_s}$ are the stellar masses determined by using three and four bands respectively. The slope is consistent with unity. To first order, the absence of the $K_{S}$-band data results in only a $\sim 0.04$ dex overestimate of the stellar mass. 

The DEEP2 galaxies with $K_{s}$-band data from the full sample are generally brighter and more massive than our selected galaxies, while the color distribution is consistent with our selected sample. The $K_{s}$-band data for our selected sample span the full range of mass, color and absolute B-band magnitude and the medians are 9.81, 0.68 and -20.7 respectively. To check for any systematic variations with galaxy properties we fit half of the data that is below the median in mass, color and magnitude respectively and find that the fits are all consistent within the errors. When $K_{s}$-band photometry is unavailable, we subtract 0.04 dex from the mass to correct for the systematic overestimate. After applying this correction, there is a 0.12 dex dispersion between the one-to-one correspondence of the two mass estimates.

We have redetermined the local MZ and LZ relations from $\sim21,000$ galaxies in the SDSS data release 7 (see appendix). We compared our stellar mass determination using Le Phare with those determined by the MPA/JHU team from fitting the photometry and a Kroupa IMF \citep{Kroupa2001}. Both methods use stellar population synthesis to fit the photometry and we expect the estimates to only vary by a constant offset.  The right panel of Figure \ref{fig:mass} shows this comparison. The dashed line is the one-to-one correspondence and the solid line is a bootstrap linear bisector fit to the data as described earlier. The linear fit is given by
\begin{equation}
X_{LP} = (0.198\pm0.001) + (1.035\pm0.003)X_{SDSS}, 
\label{eq:ke}
\end{equation}
where $X_{LP} = M_{LP} - 10$ and $X_{SDSS} = M_{SDSS} - 10$. $M_{LP}$ and $M_{SDSS}$ are the stellar masses determined using Le Phare and those provided by the MPA/JHU team,  respectively. As before, to minimize the covariance of the slope and intercept, we zero point the data by subtracting 10 from the masses. The slope is near unity and to first order the estimates differ by only a constant offset of 0.198 dex (a factor of $\sim1.6$). Taking into account this constant offset, there is a 0.14 dex dispersion in the one-to-one correspondence between the two mass estimates. 

For local galaxies it has been shown that the errors between photometric and dynamical mass are typically $\sim\!0.2$ dex \citep{Drory2004}. In this study, we observe galaxies at $z\sim0.8$ and therefore expect even greater errors in the photometric determination of mass. Moreover, \citet{Conroy2009} have shown that additional uncertainties in estimates of physical parameters from stellar population synthesis modeling result from the choice of IMF, dust model and spectral libraries. However, the full impact of these effects on the absolute calibration of the physical parameters are still not well understood. Therefore, when investigating evolution of the MZ relation, we take care to have a consistent relative calibration of the physical parameters between the samples. The absolute calibration remains uncertain.

\subsection{Correction for Stellar Balmer Aborption}

\begin{figure*}
\includegraphics[width=7in]{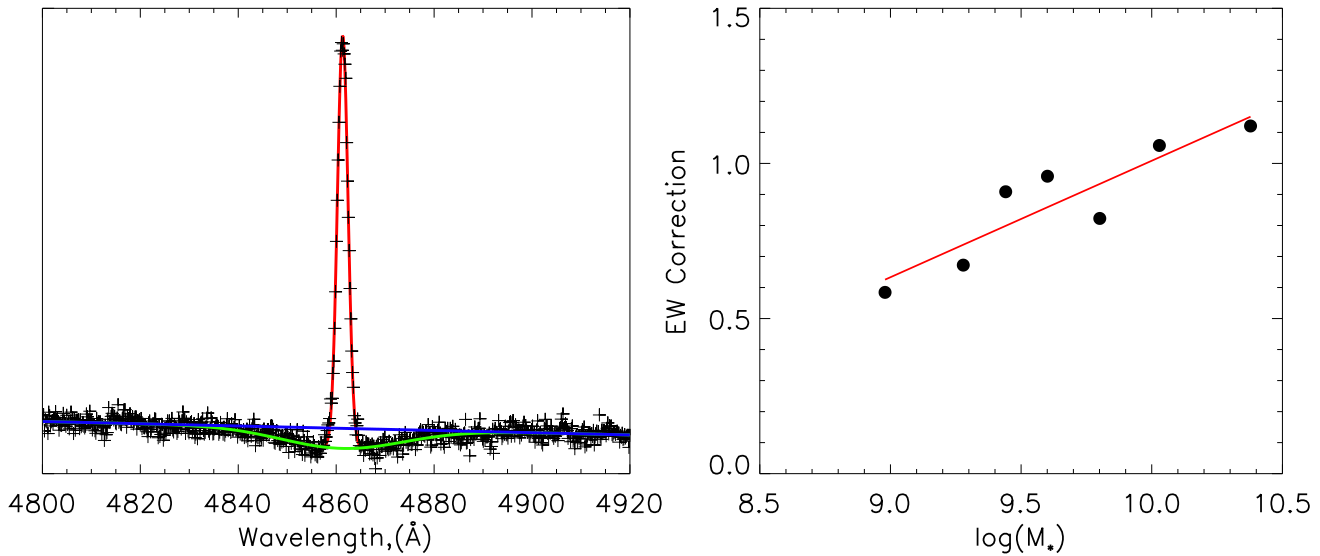}
\caption{The left panel shows an example of a stacked spectra. The red curve is a fit to the $H\beta$ emission, the green curve is a fit to the $H\beta$ absorption and the blue line is a fit to the continuum. The are of the region bracketed by the absorption on the bottom, emission on the sides and continuum at the top is the correction factor to the $EW(H\beta)$. The right panel shows this correction for $H\beta$ absorption as a function of stellar mass for data sorted into 7 mass bins. The red line is a linear parameterization of the correction as a function of stellar mass.}
\label{fig:ba}
\end{figure*}

When measuring EWs it is necessary to make a correction for underlying stellar absorption in the Balmer lines. \citet{Kobulnicky2003b} have estimated an average correction of $2\AA$ in the $EW(H\beta)$ for the underlying Balmer absorption using 22 galaxies from the spectroscopic galaxy atlas of \citet{Kennicutt1992}. The galaxies in the Kennicutt atlas have a spectral resolutions of 5-8$\AA$. We expect the effects of Balmer absorption to be greater in galaxy spectra with lower resolution because the flux of the narrow emission line is spread over a broader region of the absorption trough. Presumably for this reason, \cite{Cowie2008} find a correction of $1\AA$ by stacking spectra observed using DEIMOS onboard Keck with a spectral resolution of $3.5\AA$. The DEEP2 data have also been obtained using DEIMOS but with a smaller spectral resolution of $1.4\AA$.

Unfortunately, the low S/N in our data does not allow us to investigate the Balmer absorption in individual galaxy spectra. However, the strength of the Balmer absorption is a function of the age of the stellar population and is therefore expected to correlate with physical parameters of the galaxy such as stellar mass.  In order to investigate the effects of stellar absorption on our measurement we stack our spectra in bins of stellar mass and examine the effect of the Balmer absorption on the $EW(H\beta)$.  

We stack our spectra into 7 mass bins. We normalize each spectrum to the continuum by dividing by the median value of the continuum between $4800-4815\AA$ and $4900-4915\AA$. We then stack $\sim\!230$ spectra in each mass bin by interpolating all spectra to have the same wavelength vector and take the median of the flux corresponding to each wavelength element. We get similar results if we take the mean rather than the median. The left panel of Figure \ref{fig:ba} shows an example of a stacked spectrum. We see that the $H\beta$ emission line sits on a broad absorption trough. The $EW(H\beta)$ correction accounts for the integrated flux of the emission line (red curve) that lies below the continuum (blue curve) and would not be included when measuring the $EW(H\beta)$ in lower S/N spectra due to the absorption (green curve). To obtain a correction, we measure the $EW(H\beta)$ for the stacked spectrum and a corrected spectrum with the absorption removed.  The right panel shows the correction to the $EW(H\beta)$ as a function of stellar mass. The red line is a linear least-square fit given by
\begin{equation}
EW_{corr} = (1.01 \pm 0.05) + (0.38 \pm 0.08)x,
\label{eq:ewcorr}
\end{equation}
where $x = log(M_{\ast}) - 10$. $EW_{corr}$ is the amount added to the $EW(H\beta)$ to correct for underlying absorption. We emphasize that the correction given by equation \ref{eq:ewcorr} is sensitive to the fitting procedure. A different method (using a smaller continuum window for example) may yield a different correction. The median correction to the $EW(H\beta)$ in our sample is 0.9. We note that this correction does not significantly effect our derived MZ relation owing to the fact that the median correction of 0.9 $\AA$ is small compared to the median $EW(H\beta)$ of 12.1 $\AA$. This small correction translates to a median increase of 0.04 dex in metallicity.

We perform a simple test in order to assess the effect of varying the spectral resolution on the Balmer absorption correction. We convolve our stacked spectra with gaussians of varying widths. We find that the Balmer correction is sensitive to the spectral resolution. The correction increases with smaller spectral resolution and the slope of the correction with respect to stellar mass flattens. We speculate that the larger correction found by \citet{Kobulnicky2003b} may be attributed to the lower spectral resolution of their data. In a future study, using higher quality data we hope to quantitatively establish the magnitude of this effect.

\subsection{Metallicity Determination}

 \begin{figure}
\includegraphics[width = \columnwidth]{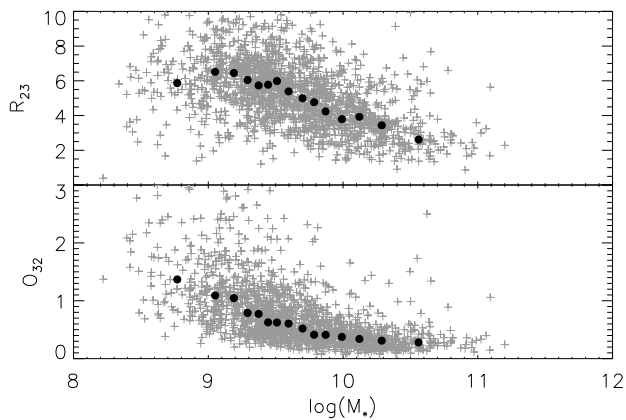}
\caption{The $R_{23}$ (top) and $O_{32}$ (bottom) line ratios as a function of stellar mass. The grey pluses are the unbinned data and the black filled circles are the data binned into 15 mass bins where each bin has equal number of data points. }
\label{fig:line_ratio}
\end{figure}

We use the strong line diagnostics of KK04 as presented in KE08 in order to obtain an estimate of galaxy gas-phase metallicities. The diagnostics are based on the \citet{Kewley2002} $R_{23}$ theoretical calibrations. In both diagnostics the metallicity is determined using the $R_{23}$ and $O_{32}$ ratios. We calculate these ratios from our EWs such that 
\begin{equation}
R_{23} = \frac{EW([OII]) + EW([OIII])}{EW(H{\beta})}
\label{eq:r23}
\end{equation}
and 
\begin{equation}
O_{32} = \frac{EW([OIII])}{EW([OII])},
\label{eq:o32}
\end{equation}
where $EW([OII])$ is for the [OII] doublet and $EW([OIII])$ is taken to be $1.33 \times EW([OIII]\lambda5007)$. We have used the assumption that the ratio of the fluxes of $[OIII]\lambda5007$ to $[OIII]\lambda4959$ is 3 \citep{Osterbrock1989}. We have fitted the $[OIII]\lambda4959$ line because the widths of the $H\beta, [OIII]\lambda4959$ and $[OIII]\lambda5007$ line fits are tied in order to increase the S/N of the fit, as described in the appendix, but the S/N in the $EW([OIII]\lambda5007)$ is higher and the $EW([OIII]\lambda4959)$ is inferred from theoretical considerations for the metallicity determination.

We use equations A4 and A6 given in the appendix of KE08 to derive the ionization parameter and metallicity respectively. The $R_{23}$ method for determining metallicities is known to be sensitive to the ionization parameter. The ionization parameter characterizes the ionization state of the gas and quantitatively represents the number of ionizing photons per second per unit area divided by the hydrogen density. The ionization parameter has the units of velocity and can be thought of as the maximum velocity of the ionization front through the nebula. The metallicity and ionization parameter are interdependent. In order to obtain a consistent measurement of the metallicity and ionization parameter an iterative scheme is used, the details of which are provided in the appendix of KE08. Furthermore, all metallicities used in this study for comparison that are not explicitly calculated using the KK04 calibration are converted to be consistent using the conversions provided in KE08.

Figure \ref{fig:line_ratio} shows the $R_{23}$ (top) and $O_{32}$ (bottom) line ratios for our sample. The grey pluses are the individual measurements and the black filled circles are the data sorted into 15 mass bins of equal number of data points. The median values of the $R_{23}$ and $O_{32}$ line ratios are 5.1 and 0.6 and median errors are 0.6 and 0.06 respectively. Both the $R_{23}$ and $O_{32}$ line ratios are sensitive to the metallicity and the relation between stellar mass and the line ratios can be attributed to the relation between stellar mass and metallicity. 

Metallicity is not a monotonic function of $R_{23}$, but rather is doubly valued for a given ratio. A particular value of $R_{23}$ corresponds to two different metallicities, one on the higher metallicity branch and one on the lower metallicity branch. The peak of the $R_{23}$ ratio occurs at $12 + log(O/H)\approx8.4$. This degeneracy is due to the fact that on the lower branch $R_{23}$ scales with metallicity because the intensity of the collisionally excited [OII] and [OIII] lines scales with the abundance. On the upper branch, nebular cooling, which results from collisional excitation followed by photon emission, effectively cools the nebula decreasing the electron temperature leading to a decrease in the rate of collisional excitation of the [OII] and [OIII] lines. In order to break this degeneracy, generally other line ratios such as [NII]/$H_{\alpha}$ are used. However, in our sample, these lines are not observed and the metallicity is assumed to lie on the upper branch. This assumption breaks down at lower masses. Looking at the $R_{23}$ mass relation in Figure \ref{fig:line_ratio}, a turnover appears to occur at about $log(M_{\ast}) \sim 9.2$ and we revisit this issue in $\S \,4$.

 \section{Mass-Metallicity and Luminosity-Metallicity Relations}
 
 \begin{figure}
\includegraphics[width = \columnwidth]{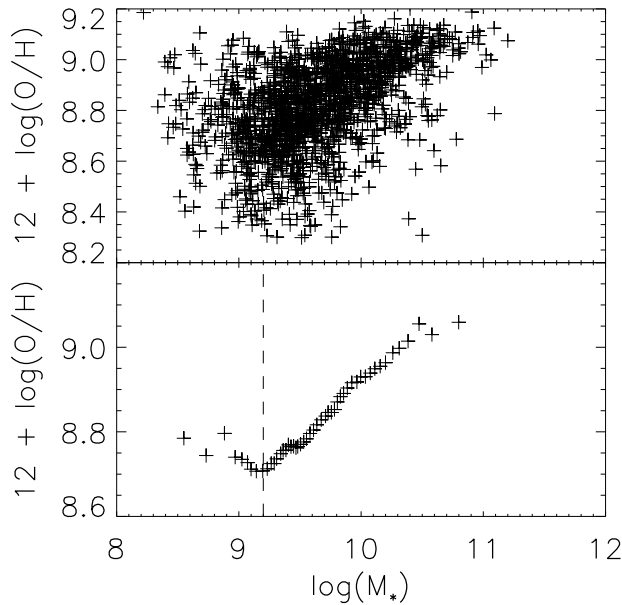}
\caption{The mass and metallicity for the DEEP2 sample. The top panel shows the metallicity plotted against mass for the 1,631 galaxies in our sample. The bottom panel is a boxcar averaged smoothing of the data displayed in the top panel binned by mass. We attribute the break in the slope at $M\sim10^{9.2}M_{\odot}$ to the misplacement of galaxies on the upper branch of the KK04 metallicity diagnostic. The vertical dashed line $(M = 10^{9.2}M_{\odot})$ in the lower panel is our lower limit cutoff mass used in determining the MZ relation for the DEEP2 sample.}
\label{fig:low}
\end{figure}

From our measured EWs and photometry, we can estimate both the  gas-phase oxygen abundances and stellar masses for the galaxies in our sample. From the Spearman rank test, we conclude with over 99\% confidence that we have a positive correlation between mass and metallicity and luminosity and metallicity. This confidence in our observed correlation justifies a fit to the data. 
 
 \subsection{The Mass-Metallicity Relation}

 \begin{figure*}
\includegraphics{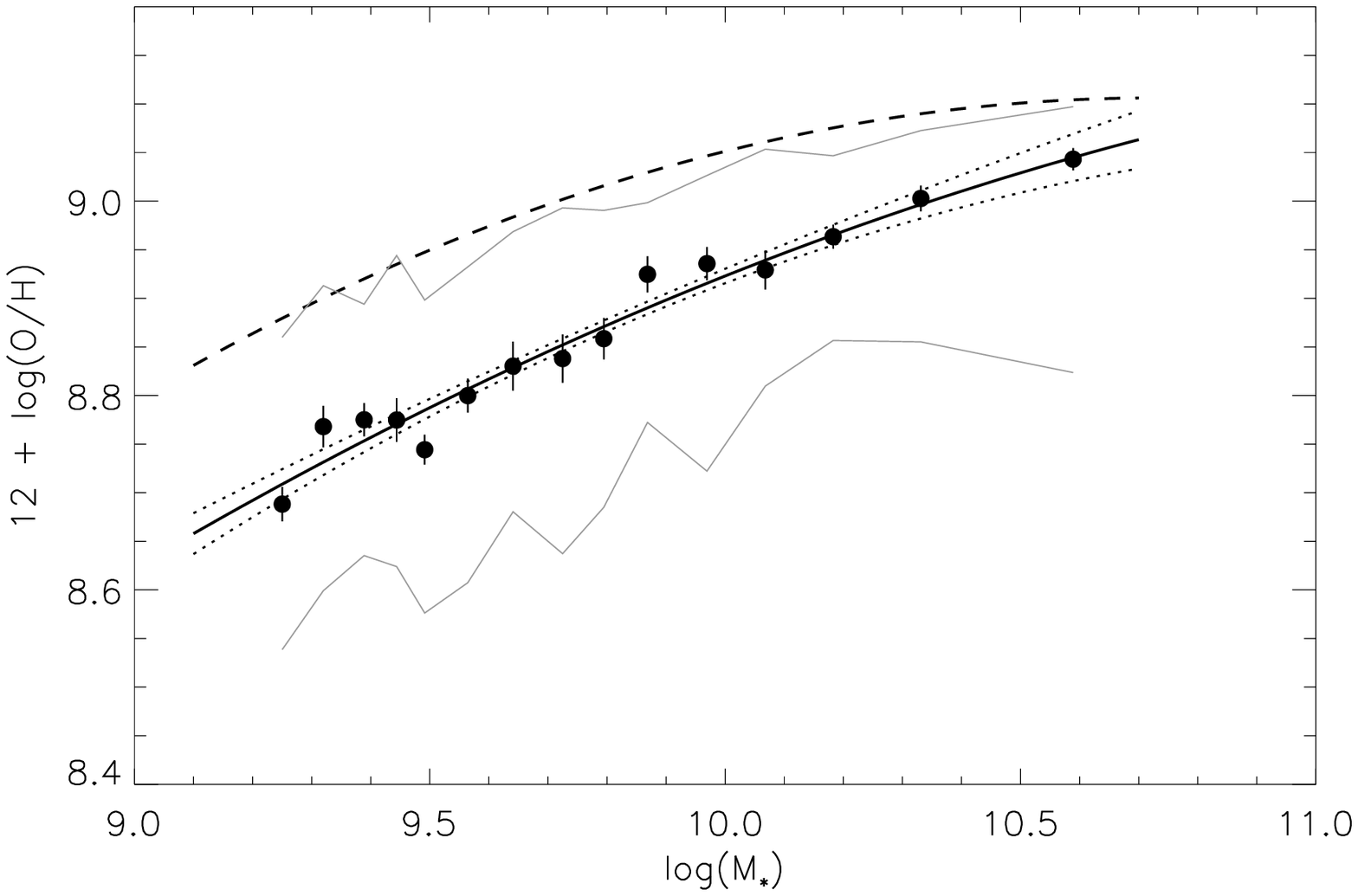}
\caption{The MZ relation derived from the DEEP2 sample. The black data points are the median of the mass and metallicity sorted into 15 bins. The bootstrapped errors are determined by randomly selecting galaxies from within each mass bin and determining the median for this subsample. The error bar then is the standard deviation of the distribution of median values. The grey lines are the $16\%$ and $84\%$ contours of the data. The solid curve is a quadratic fit to the data and the dotted curves around the quadratic fit are the $1\sigma$ uncertainty in the fit determined from bootstrapping the errors. The dashed curve is the local MZ relation from the appendix of this paper.}
\label{fig:mz}
\end{figure*}
 
The top panel of Figure \ref{fig:low} shows stellar mass and metallicity measured for our sample of 1,631 galaxies. The bottom panel shows binned data with a boxcar smoothing. In determining the metallicity, we have assumed that all galaxies fall on the upper branch of the KK04 metallicity diagnostic. T04 have shown that in the local universe the MZ relation extends down to a stellar mass of $M_\ast=10^{8.5}M_\odot$ with no break in the slope. We work under the assumption that the MZ relation at $z\sim0.8$ extends to at least $M_\ast \sim 10^{9}M_\odot$ with no break in slope. Under this assumption, we interpret the break in the slope at $M_\ast = 10^{9.2}M_\odot$ observed in the bottom panel of Figure \ref{fig:low} to be caused by the misplacement of lower branch galaxies on the upper branch. This turnover mass is further evidenced by the the $R_{23}$ ratio as a function of stellar mass seen Figure \ref{fig:line_ratio}. We take a lower mass limit, indicated by the vertical dashed line in the bottom panel of Figure \ref{fig:low}, of $M_\ast=10^{9.2}M_{\odot}$ to fit our MZ relation. We note that 283 of our 1,631 galaxies fall below this limit, leaving us with 1,348 galaxies from which we determine the MZ relation.

We parameterize the MZ relation with a quadratic function of the form 
\begin{equation}
12 + log(O/H) = A + Bx + Cx^2,
\label{eq:quad} 
\end{equation}
where $x = log(M_{\ast}) - 10$. Our data and fit residuals are not normally distributed so we apply a standard non-parametric bootstrapping technique in determining the fit parameters and errors. One of the underlying assumptions of the bootstrapping method is that the observed distribution reasonably approximates the parent distribution. Under this assumption and the assumption of independence of the sample, properties determined from many independent random subsamples of the data should be normally distributed. We begin by randomly selecting, with replacement, 1,348 galaxies from our sample of 1,348 galaxies. Meaning, within a random subsample some data points may be selected more than once while others may not be selected at all. We then bin these randomly selected galaxies by sorting them into 15 equally populated mass bins. The mass and metallicity in each bin is determined by taking the median of the data in the bin. We perform a least-square quadratic fit to the binned data using the routine $\emph{poly\_fit.pro}$ in IDL. We randomly resample and fit the data 10,000 times and take the mean and standard deviation of this distribution of fitted parameters, which are normally distributed, as the fit parameters and errors for the MZ relation.

The MZ relation is best fitted by
\begin{equation}
12 + log(O/H) = 8.923 + 0.24 \, x - 0.06 \, x^2,
\label{eq:mzfit}
\end{equation}
with the $1\sigma$ errors in each of the parameters in equation \ref{eq:mzfit} given by $\sigma_{A} = 0.007, \sigma_{B} = 0.01$ and $\sigma_{C} = 0.03$. Figure \ref{fig:mz} shows the DEEP2 MZ relation at $z\sim0.8$ compared to the local MZ relation from SDSS (see appendix). The black filled circles are the data sorted into 15 mass bins. Each bin contains $\sim 90$ data points and the errors on the data are determined by randomly selecting $\sim90$ data points in each bin and taking the median of the subsample. We take the standard deviation of this distribution as the error in each bin. We note that this scatter characterizes the distribution of metallicities within each mass bin and therefore accounts for both measurement uncertainties and intrinsic scatter in the metallicity determination. The grey lines are the $16\%$ and $84\%$ percentile contours of the data. The large dispersion at the high mass end is due to the large interval in mass covered by the highest mass bins.

Overall, we find that the high-mass ($M > 10^{10.5} M_\odot$) DEEP2 galaxies have similar metallicities (within the errors of $\pm 0.05$~dex) to local galaxies at the same stellar mass.  At lower masses ($M < 10^{10.5} M_\odot$), the DEEP2 galaxies have lower metallicities than the SDSS sample for a given stellar mass, up to a $\Delta log(O/H) \sim 0.15$ dex.

\subsection{The Luminosity-Metallicity Relation}

\begin{figure*}
\includegraphics{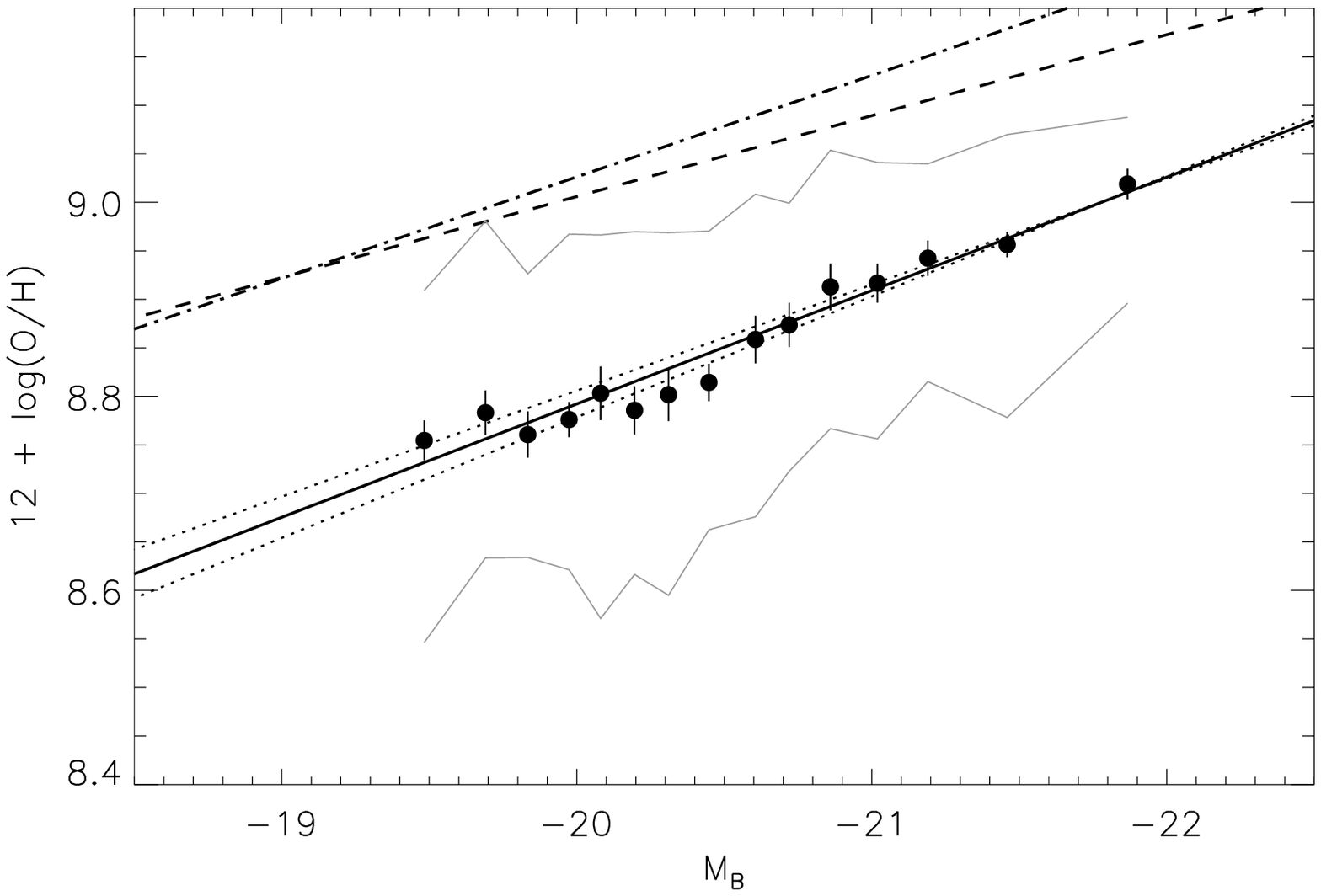}
\caption{The LZ relation derived from the DEEP2 sample. The black data points are the median of the $M_{B}$ and metallicity sorted into 15 luminosity bins. The solid line is a linear fit to the data and the dotted lines are the $1\sigma$ uncertainty of the fit determined from bootstrapping errors. The grey lines are the $16\%$ and $84\%$ percentile contours of the data. The dashed line is the local LZ relation fit over the full range of magnitudes ($-21<M_{B}<-18$). The local LZ relation appears to turn over for brighter galaxies and the dot-dashed line is a fit to the linear part of the relation ($M_{B}> -20$). See appendix of this paper for more details.}
\label{fig:lz}
\end{figure*}

Figure \ref{fig:lz} shows the LZ relation derived from the DEEP2 sample and the local sample from SDSS. Similar to the MZ relation, the LZ relation is determined for galaxies with $M_{\ast} > 10^{9.2}M_{\odot}$. The filled circles are the median metallicities in 15 luminosity bins. The solid curve is a linear fit to the binned data and is parameterized by the equation
\begin{equation}
12 + log(O/H) =  (8.909 \pm 0.006) - (0.117 \pm 0.008)X_{B},
\label{eq:lin}
\end{equation}
where $X_B = M_B + 21$. The errors in the fit and binned data of the LZ relation are derived in the same manner as for the MZ relation. The grey lines are the $16\%$ and $84\%$ percentile contours of the data. The dashed curve is the local LZ relation from SDSS taken from the appendix of this paper. The fit made is over the full range of magnitudes ($-21<M_{B}<-18$) of the local LZ relation. However, the local LZ relation appears to turn over at $M_{B}<-20$. The dot-dashed curve is a fit to the linear part of the local LZ relation with $M_{B}> -20$. The slope of this fit and the DEEP2 LZ relation are consistent to within the errors. KK04 find that the slope of the LZ relation for $z = 0.6-0.8$ and $z = 0.8-1.0$ are $-0.117\pm0.017$ and $-0.134\pm0.032$ respectively. To within the errors, the slopes derived by KK04 agree with DEEP2 and local LZ relation slopes as well.

%The lack of evolution observed in the LZ slope suggests that the B-band magnitude is an excellent indicator of the relative metallicities of a sample of blue emission line galaxies at the same redshift in the interval of $z=0-1$. Given a sample of star-forming galaxies, the relative metallicities can be estimated from their absolute B-band (AB) magnitude by
%\begin{equation}
%\Delta(log(O/H) = -(0.121 \pm 0.008) \Delta M_{B}
%\label{eq:mb}
%\end{equation}
%where the $\Delta(log(O/H)$ and $\Delta M_{B}$ are the differences in metallicity and absolute B-band magnitude respectively. The slope and error are determined from the average of the DEEP2 and local LZ slopes and rms of the errors respectively. The relative metallicities in individual galaxies cannot be measured due to the intrinsic dispersion in the LZ relation. The absolute metallicities cannot be estimated due to B-band luminosity evolution between $z = 0-1$ as discussed in the following section. 

\section{Discussion}

In this section we will discuss our key assumptions ($\S \, 5.1$), the evolution of the B-band luminosity as a function of stellar mass in the population of blue galaxies ($\S \, 5.2$) and compare our determination of the MZ relation with other authors ($\S \, 5.3$).

\subsection{Assumptions}

We list and address some key assumptions made in this paper:

\vspace{.2 cm}

1. All galaxies with $M_\ast \geq 10^{9.2} M_\odot$ are on the upper branch of the $R_{23}$ diagnostic. At the lower mass end of the DEEP2 MZ relation the dispersion is small and consistent with the dispersion found in the local MZ relation (T04). This small dispersion suggests that at $z\sim0.8$, the vast majority of galaxies with $M_\ast > 10^{9.2} M_\odot$ are on the upper metallicity branch of the $R_{23}$ diagnostic. Additionally, T04 have observed a continuous slope down to $M_\ast \sim 10^{8.5} M_\odot$. A breakdown of this assumption would result in the $R_{23}$ measure of metallicity in low metallicity galaxies to be significantly overestimated ($\sim 1$dex) resulting in a flattening of the MZ relation at lower masses. This flattening is observed to occur at $M_{\ast} = 10^{9.2}M_\odot$ and given the small dispersion, we hypothesize that misplacement on the upper branch is minor above this mass.  Near infrared spectroscopy to obtain additional line ratios for a representative subsample of DEEP2 galaxies is needed to test this hypothesis. 
\vspace{.2 cm}

2. Metals are instantaneously recycled. This approximation was first introduced by \citet{Schmidt1963}. The approximation is valid when metals produced in massive stars, such as oxygen, are considered and if the SFR is not subject to extreme variations on short timescale \citep{Pagel1997}. From their study of DEEP2 galaxies since $z=1.1$, \citet{Noeske2007a} determine that the dominant form of evolution is a gradual decline in star formation. 
\vspace{.2 cm}

3. We have attempted to remove AGN using empirical limits on the $R_{23}$ ratio and the color bimodality. This removes most of the galaxies whose emission is dominated by AGN, but composite galaxies in our sample may be blue and not exceed the empirical $R_{23}$ limit. AGN contamination in composite galaxy spectra increases the $R_{23}$ ratio due to the high $[OIII]/H\beta$ ratio. On the upper metallicity branch of the $R_{23}$ diagnostic, the metallicity is a monotonically decreasing function of $R_{23}$. Therefore, AGN contamination will lower the metallicity estimate. However, due to their relatively small numbers in the DEEP2 blue galaxy sample \citep{Weiner2007} and the small dispersion in the MZ relation, AGN contamination is  likely to play a minor role in the MZ relation \citep{Lamareille2009}.  Near infrared spectroscopy to obtain additional AGN-sensitive emission-line ratios is required to verify this assumption.

%We illustrate the effect of AGN contamination on the derived MZ relation in figure \ref{fig:agn}. We redetermine the MZ relation but remove 10 (red curve) and 20\% (blue curve) of the lowest metallicity data in each bin before applying the procedure described in $\S \, 4.1$. The redetermined MZ relation with 10 and 20\% of the data removed varies by 0.02 and 0.04 dex respectively at $M=10^{9.7} M_{\odot}$ from the MZ relation given by equation \ref{eq:fit}. Due to the AGN contamination, the reported MZ relation fit given in equation \ref{eq:fit} is a lower-limit. Figure \ref{fig:agn} shows that even removing a substantial fraction of the lowest metallicity data does not significantly alter the determined MZ relation. AGN contamination does not appear to be a significant source of error and therefore does not  alter the results and conclusions of this study.
\vspace{.2 cm}

4. Since its introduction by \citet{Salpeter1955}, the initial stellar mass function (IMF) - describing the mass distribution of stars at birth - has largely been taken to be universal and invariant. Indirect evidences for variations in the local IMF have been provided by the comparison of $H\alpha$ emission to the far-ultraviolet flux \citep{Meurer2009} and the $H\alpha$ flux to broadband color index \citep{Hoversten2008}. Comparisons of the rate of luminosity evolution to that of color evolution provides indirect observational evidence for a possible redshift evolution in the IMF \citep{vanDokkum2008}. Furthermore, IMF variations have been cited as the possible cause of the MZ relation \citep{Koppen2007}.  We work under the assumption that the IMF is universal.

\vspace{.2 cm}

5. MZ and LZ relation studies assume that the absence of low surface brightness galaxies (LSBGs) do not significantly bias the determination of the MZ and LZ relations. \citet{Impey1997} point out that LSBGs will be absent in magnitude limited surveys. However, by comparing photometry in overlapping fields where both CFHT and Hubble Space Telescope imaging is available, no galaxies brighter than the magnitude limit of $R_{AB} = 24.1$ are found to be lost due to low surface brightness \citep{Simard2002, Willmer2006}. \citet{Melbourne2007} come to similar conclusions regarding the loss of low surface brightness galaxies from deep imaging in the GOODS-N field. 

\subsection{B-band Luminosity Evolution}

In many previous studies, luminosity has been taken as a proxy for mass \citep{Kobulnicky2000, Pettini2001, Garnett2002, Kobulnicky2003a, Kobulnicky2004, Maier2004, Shapley2004, Shapley2005}. However, it has been recognized that the evolution in the LZ relation cannot straightforwardly be attributed to metallicity evolution, but that luminosity evolution must also be considered. By comparing figures \ref{fig:mz} and \ref{fig:lz}, we see that the offset in metallicity in the MZ relation is significantly smaller than the offset in the LZ relation. \citet{Lamareille2009} (among others) suggest that the differential evolution in the LZ and MZ relation results from an evolution in the mass-to-light ratio. They estimate the evolution such that
\begin{equation}
\Delta M_{B} = a^{-1} \times [\Delta log(O/H)^M - \Delta log(O/H)^L],
\label{eq:lzevol}
\end{equation}
where $a$ is the slope of the LZ relation and $\Delta log(O/H)^M$ and  $\Delta log(O/H)^L$ are the differences in metallicity at a fixed stellar mass and fixed B-band magnitude in the MZ and LZ relations, respectively. Equation \ref{eq:lzevol} assumes that the metallicity evolution is characterized by the MZ relation and that additional evolution inferred from the LZ relation is due to evolution of the luminosity. 

 \begin{figure}
\includegraphics[width = \columnwidth]{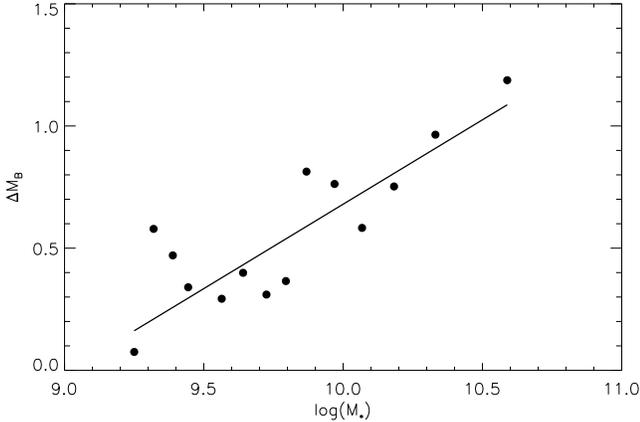}
\caption{The B-band luminosity evolution inferred from the comparison of our MZ and LZ relations. The black data points are determined by taking the difference between the binned DEEP2 data and the fit to the local MZ relation and plugging it into equation \ref{eq:lzevol}. The line is from combining equations \ref{eq:lzevol} and \ref{eq:dm}.}
\label{fig:lz_evol}
\end{figure}

We estimate the evolution of the B-band luminosity as a function of stellar mass from equation \ref{eq:lzevol}. We take the difference between the local and DEEP2 LZ relation, $\Delta log(O/H)^L$, to be the average difference between the binned DEEP2 data and the fit to the local LZ relation over all magnitudes (see Figure \ref{fig:lz}). Consequently, $\Delta log(O/H)^L = 0.20 \pm 0.03$ dex and $a = -0.11$. We take $\Delta log(O/H)^M$ to be the difference between the DEEP2 binned data and local MZ relation fit. The difference is parameterized by 
\begin{equation}
\Delta log(O/H)^M = (0.12\pm0.01) - (0.08\pm0.02)x,
\label{eq:dm}
\end{equation}
where $x = log(M_{\ast}) - 10$. Here we have performed a linear least-squares fit using $\emph{poly\_fit.pro}$ and the errors are determined from the residuals of the fit and do not account for observational uncertainties. 

Figure \ref{fig:lz_evol} shows the B-band luminosity evolution inferred from our comparison of the DEEP2 MZ and LZ relations. Combining equation \ref{eq:lzevol} and \ref{eq:dm} we get
\begin{equation}
\Delta M_{B} = (0.68\pm0.06) + (0.7\pm0.1)x,
\label{eq:bm}
\end{equation}
the B-band luminosity evolution of blue star forming galaxies between $z\sim0.8$ and the local galaxy population as a function of stellar mass. This is shown in Figure \ref{fig:lz_evol} by the solid line.

It has been shown that the luminosity functions of blue galaxies evolve with time \citep{Ilbert2005, Blanton2006, Willmer2006, Faber2007}.  These studies conclude that $M_B^\ast$ has dimmed by $\sim1$ mag since $z\sim0.8$, consistent with the evolution observed in our sample at the higher stellar mass end. A differential evolution in the B-band Tully-Fisher relation has also been observed since $z\sim1$ and is consistent with the observed luminosity evolution \citep{Weiner2006}. \citet{Noeske2007a} show that the star formation as a whole was higher at $z\sim1$ and that the star formation rate gradually declines with galaxies spending 67\% of their lifetime since $z = 1$ with SFR that are within a factor 2 of their average and 95\% of their time within a factor of 4. In an accompanying letter, \citet{Noeske2007b} find that initial onset and rate of decline in star formation are functions of redshift and galaxy mass with less massive galaxies having later initial onsets with a lower rate of decline and consider this an important component of downsizing \citep{Cowie1996}. The absolute B-band luminosity is a tracer of star formation and the observed differential evolution with respect to stellar mass shown in Figure \ref{fig:lz_evol} may be attributed to this type of downsizing.

%\subsection{Evolution of LZ Slope}

%KK04 measure the LZ relation in four redshift intervals using 204 galaxies from the Team Keck Redshift Survey in the redshift interval of $0.2<z<1$. The observe that zero point of the linear fit to the LZ relation decreases with redshift, meaning that galaxies at a given luminosity become more metal poor with increasing redshift. The observe a possible evolution in the slope of the LZ relation such that the slope of the relation steepens with increasing redshift. This suggests that the least luminous galaxies become increasingly metal poor with increasing redshift and they interpret this to be consistent with the downsizing scenario.

\subsection{Comparison of MZ Relations}

 \begin{figure}
\includegraphics[width = \columnwidth]{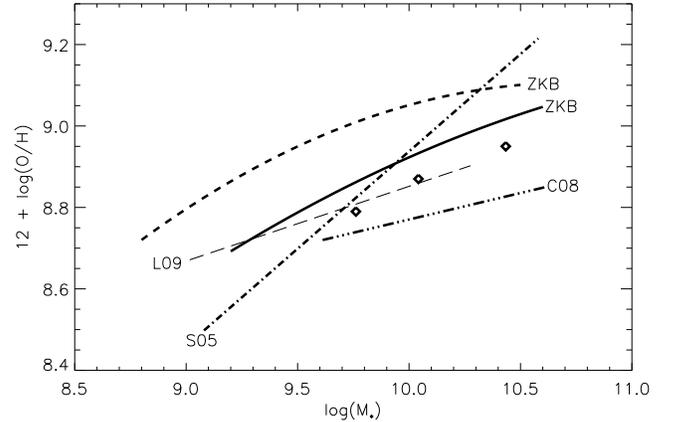}
\caption{A comparison of the MZ relation derived by several authors. The solid line in this figure is our MZ relation derived for galaxies with $z = 0.75-0.82$ (labeled ZKB). The short dashed curve is the local MZ relation for galaxies with $z = 0.04-0.1$ from the appendix of this paper (labeled ZKB). The long dashed curve is the MZ relation for galaxies with $z = 0.6 - 0.8$ \citep[][labeled L09]{Lamareille2009}. The dot-dashed curve is for galaxies with $z = 0.4 - 0.98$ \citep[][labeled S05]{Savaglio2005}. The triple dot-dashed curve is for galaxies with $z = 0.475 - 0.9$ \citep[][labeled C08]{Cowie2008}. The diamonds are the median instead of the mean of the data from \citet{Cowie2008}. Metallicities and masses have been converted to be consistent amongst the samples. The relations only cover the stellar mass range over which they were determined.} 
\label{fig:mz_compare}
\end{figure}

There are considerable difficulties in making a comparison of the MZ relation owing to the varying quality and type of data and the different methodologies and sample selections. The systematic differences among strong-line metallicity calibrations have been investigated by KE08.  They derived conversions that reach agreement among calibrations to within $\pm 0.05$~dex in metallicity.  KE08 conclude that while absolute metallicity values are uncertain, relative metallicities (such as MZ relations) can be reliably compared, providing that the metallicities have been converted into the same base calibration.

Though offsets between different metallicity calibrations can now be removed, the systematic effects on differing measurements of stellar mass have yet to be fully explored.  Moreover, comparisons between spectral and photometric methods have not been rigorously investigated and the systematic effects of the photometric bands used in determining stellar mass remain uncertain. These problems are further compounded by the fact that many of the selection criteria in these studies may be a function of redshift (i.e. S/N of flux or EW, minimum values of flux or EW, color, etc.) and therefore the samples from which the MZ relations are determined may not be comparable even if similar selections are used.

Several authors have investigated the MZ relation covering the redshift interval examined in this study ($z = 0.75 - 0.82$). Figure \ref{fig:mz_compare} compares our MZ relation (solid curve) with previously derived relations from \citet{Savaglio2005} (dot-dashed), \citet{Cowie2008} (triple dot-dashed) and \citet{Lamareille2009} (long dashed line). All metallicities have been converted to the KK04 calibration using the conversions in KE08. We have attempted to make the stellar masses consistent with ours by converting all the determinations to the Chabrier IMF used in this study. The relations plotted only cover the stellar mass range over which they were determined. 

The first attempt to investigate the MZ relation at intermediate redshifts was made by \citet{Savaglio2005} using the Gemini Deep Deep Survey. This sample consists of 28 galaxies from the Gemini Deep Deep Survey with emission in [OII], $H\beta$ and [OIII] at the $3\sigma$ level covering the redshift range of $z = 0.40 - 0.98$. This sample consists of galaxies designated as late (7/28) and intermediate type (21/28). This sample is augmented with 28 galaxies from the Canada France Redshift Survey \citep{Lilly1995}.The dot-dashed fit in Figure \ref{fig:mz_compare} is a linear bisector fit of the S05 mass and metallicity. Though Savaglio et al. establish a MZ relation at higher redshifts, the small sample size and selection in addition to the fitting method does not allow us to make a direct comparison with our determination.

\citet{Lamareille2009} study the evolution of the MZ relation by splitting their sample into three redshift ranges. They find that the most massive galaxies undergo the most evolution at $z\sim0.8$, implying that the slope of the MZ relation flattens out at higher redshfits. However, L09 do not find a strong Spearman rank correlation coefficient, thus hindering a rigorous statistical analysis. The relation shown in Figure \ref{fig:mz_compare} (long-dashed line) is a linear fit to  average metallicities in 4 mass bins for $\sim130$ galaxies in the redshift range of $z = 0.6 - 0.8$. The galaxies are selected to have $S/N > 4$ in the flux of the [OII], $H\beta$ and [OIII] emission lines. The flattening of the slope at higher redshifts may possibly be a consequence of the S/N cut on the [OII] and [OIII] lines. These cuts may be a function of redshift and bias against higher metallicity galaxies generally found at higher stellar masses. At lower masses, the MZ relation of L09 is in agreement with our determination. 

\citet[][hereafter C08]{Cowie2008} study the evolution of the MZ relation by splitting their sample into a low-z and mid-z redshift interval. The two intervals are $0.05<z<0.475$ and $0.475<z<0.9$ and have a median redshift of 0.44 and 0.75 respectively. The low-z and mid-z samples have 35 and 154 galaxies respectively. They find that there is a 0.25~dex difference in the metallicity in the mid-z sample as compared with the local relation and that the metallicity increases by 0.13 dex between the mid-z and low-z sample. The triple dot-dashed line in Figure \ref{fig:mz_compare} shows the mid-z relation. 

A major source of discrepancy between our determination and C08 can be attributed to the fitting procedure. Figure 31 of C08 shows that the spread in metallicity becomes larger as metallicity decreases. This may result from the greater uncertainty in determining metallicity from the $R_{23}$ diagnostic which becomes less sensitive to metallicity around the $R_{23}$ local maximum ($12 + log(O/H)\sim8.4$) (see C08 figure 23 or KK04 figure 7 for example). The relation is curved so symmetric errors in $R_{23}$ become asymmetric errors in metallicity. In a least-square fit the data are assumed to be normally distributed, however in this case they are not and outliers exert greater influence in determining the best fit. Looking at figure 63 of C08 shows that the median of the metallicity in 4 mass bins is systematically higher than the fitted relation. With the diamonds in figure \ref{fig:mz_compare}, we plot the median of the MZ relation for the mid-z sample sorted into 3 mass bins (L. Cowie, private communication). We fit our relation to the median metallicity in 15 mass bins as described in $\S \, 4$.  The MZ relation determined from the median of the C08 data is more comparable to our determination. Given the many uncertainties discussed in comparing the MZ relation, the agreement between the relation of L09, C08 and our determination from DEEP2 is promising. 

Our MZ relation comparisons highlight the need for consistent sample selection and MZ relation fitting techniques when comparing MZ relations at the same or different redshifts.

\section{Summary}

We have conducted a study of the mass-metallicity and luminosity-metallicity relation at $z\sim0.8$ using the 1,348 galaxies from the DEEP2 survey.  This large sample has allowed us to establish the mass-metallicity and luminosity-metallicity relation at $z\sim0.8$ with greater statistical significance than previous studies in this redshift regime.  Such a large sample within a small redshift range is essential for avoiding evolution effects in luminosity or metallicity that may influence luminosity-metallicity and mass-metallicity relations derived over a larger redshift interval.

We have obtained stellar masses by inferring a SED from our photometry and fitting it with current stellar population synthesis models. We have determined the gas-phase oxygen abundance in a consistent manner from strong-line methods. We summarize our main results as follows:
\vspace{.5 cm}

1. We find a clear correlation between mass and metallicity at $z\sim0.8$, consistent with previous work \citep{Savaglio2005, Cowie2008, Lamareille2009}. The level of chemical enrichment achieved by galaxies is a function of stellar mass. At $z\sim0.8$, galaxies with $M_\ast\sim10^{10.6} M_{\odot}$ have metallicities that are lower by 0.05 dex as compared to the local sample, which is within the errors of the metallicity calibration conversions.  
The metallicity difference between $z\sim0.8$ and local galaxies rises to 0.15 dex at $M_\ast\sim10^{9.2} M_{\odot}$.

\vspace{.2 cm}

2. There is a clear correlation between luminosity and metallicity at $z\sim0.8$, as shown in previous work. The level of chemical enrichment achieved by galaxies is a function of luminosity.  The metallicity at a given luminosity at $z\sim0.8$ is 0.20 dex lower than the local sample due to the combination of metallicity and luminosity evolution with redshift.

\vspace{.2 cm}

3. The offset in the LZ relation between $z\sim0.8$ and local galaxies is significantly larger than the offset observed in the MZ relation between these two redshifts.  We attribute this difference to evolution in the mass-to-light ratio between $z\sim0.8$ and $z\sim0$.  We infer a luminosity evolution that scales with stellar mass and is consistent with determinations from studies of blue galaxy luminosity functions.

\vspace{.5 cm}

Both the luminosity and metallicity evolution are a function of stellar mass, consistent with the downsizing scenario of galaxy evolution.  In a subsequent paper, we will examine the origin of the mass-metallicity relation and its evolution over cosmological timescales.

%\vspace{3 mm}
\acknowledgments
We thank the referee Ben Weiner for his careful reading and many useful suggestions for improving the paper. L. Kewley and J. Zahid gratefully acknowledge support by NSF EARLY CAREER AWARD AST07-48559.  We thank the DEEP2 team for making their data publicly available. The analysis pipeline used to reduce the DEIMOS data was developed at UC Berkeley with support from NSF grant AST-0071048. We also thank Kevin Bundy for generously sharing his K-band photometry and Christy Tremonti for making her measurements available. We are grateful to Stephane Arnout and Olivier Ilbert for making their photo-z code available for use in estimating galaxy stellar mass and are thankful to Olivier Ilbert and C.J. Ma for help in installing and implementing Le Phare. Finally, we would like to thank Charlie Conroy, Christy Tremonti, Kristian Finlator, Len Cowie, David Rupke, Dave Sanders, Chiaki Kobayashi, Ezequiel Treister and TC (the player) and Josh Barnes for useful discussion. We acknowledge the cultural significance Mauna Kea has for the Hawaiian community and with all due respect say mahalo for its use in this work.

\appendix
\subsection{Line fitting and line ratio determination}
Conventionally, when spectra are not flux calibrated, as is the case in most spectroscopic redshift surveys, equivalent widths are utilized in determining the emission line ratios used for inferring metallicity \citep[hereafter KP03]{Kobulnicky2003b}. The essential feature of equivalent widths is the normalization of the line emission to the underlying continuum. Here we will discuss the algorithm developed to determine line ratios in a self-consistant manner.

%\begin{figure*}
%\includegraphics{fit_scheme.eps}
%\caption{The four panels illustrate the sequence of our fitting scheme for emission lines. Panel A shows the flux vector of a typical spectra, $F(x)$, plotted against the wavelength vector, $x$, corresponding to each resolution element. The red line is a fit of the continuum, $C(x)$, to the spectra. In panel B $N(x)$ is the flux, $F(x)$, normalized to the global continuum, $C(x)$. A three parameter gaussian fit is overplotted in red. The dashed lines show the $\pm10\sigma_{e}$ of line center, $x_{e}$, for determining the local flux. Panel C shows the local flux, $L(x)$. The blue curve is a 6-degree polynomial, $p(x)$, modeling the underlying local continuum. The red curve is a gaussian modeling the emission line. $R(x)$, panel D, is the local flux, $L(x)$, normalized to the local continuum, $p(x)$. The red curve is the fitted gaussian from which the pseudo-EW of the line is calculated. We subtract the fitted gaussian from $R(x)$ and take the rms of the data to be the noise in the continuum, $\sigma_{e}$.}
%\label{fig:fit}
%\end{figure*}

We define $F(x)$ as the flux vector of our spectra with $x$ being the rest-frame wavelength vector corresponding to each resolution element. We mask out $\pm 5 \sigma_{v}$ of our rest-frame line center wavelengths (from NIST\footnote{National Institute of Standards and Technologies}) and fit a global continuum. Here we take $\sigma_{v}$ to be a velocity dispersion of $150$ $km/s$. We model the global continuum as
\begin{equation}
\eqnum{1A}
C(x) = \displaystyle\sum_{i=1}^{39} c_{i} T_{i}(x) + \displaystyle\sum_{j=0}^2 p_{j}x^{j}
\end{equation}
where $T_{i}(x)$ are model stellar spectra obtained from \citet{Bruzual2003}. We perform a bounded value non-linear least square fit using the MPFIT\footnote{http://purl.com/net/mpfit} set of routines \citep{Markwardt2009} in IDL to obtain $c_{i}$ and $p_{j}$ with the constraint that $c_{i}\geq0$ . We obtain a continuum-normalized flux, such that 
\begin{equation}
\eqnum{2A}
N(x) = \frac{F(x)}{C(x)} - 1,
\end{equation}
and fit a three parameter gaussian, $A_{N} e^{- \frac{(x-x_{N})^2}{2 \sigma_{N}}}$ to all emission lines.  $A_{N}, x_{N}$ and $\sigma_{N}$ are the gain, line center and sigma of the gaussian and are used as initial estimates for fits to the local flux. In all our fits to the DEEP2 data, we derive the errors on our parameters by propagating the measurement uncertainties in the spectrum.

\begin{figure}[h]
\begin{center}
\includegraphics[width=.7\columnwidth]{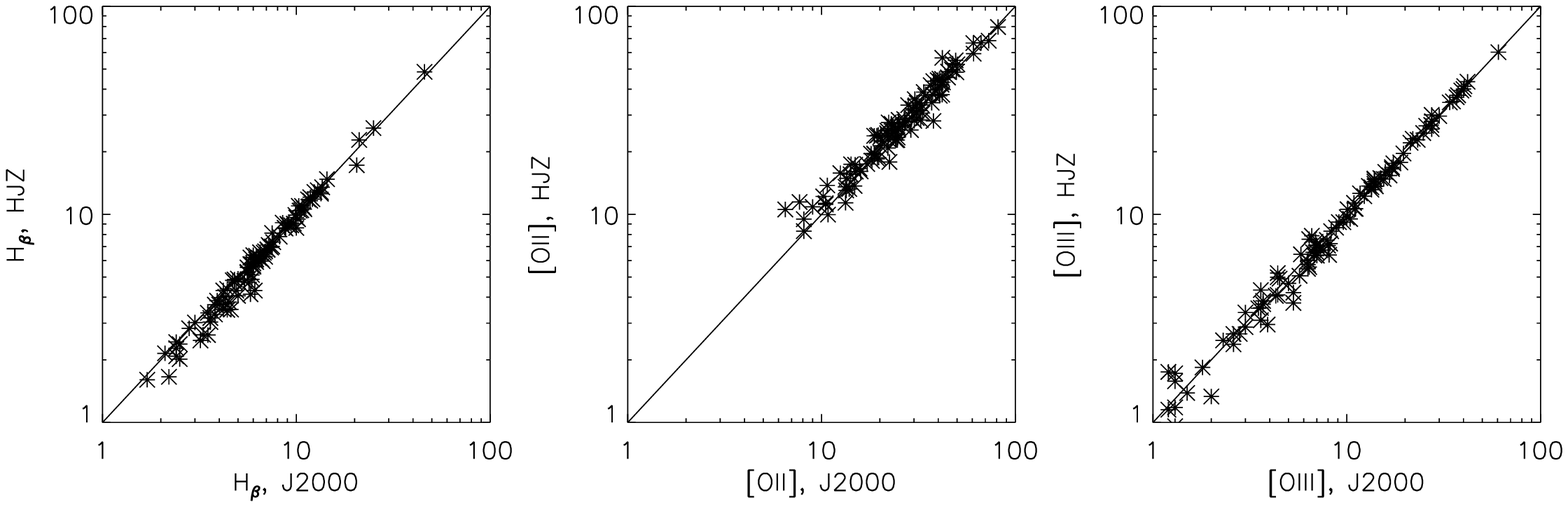}
\end{center}
\caption{Comparison of fitted emission-line pseudo-EWs (denoted 'HJZ', y axis) to the reported equivalent widths (denoted 'J2000', x axis) for 102 galaxies from the Nearby Field Galaxies Survey \citep{Jansen2000}. From left to right the plots show the $H_{\beta}$, [OII] and [OIII] comparisons respectively. The solid line in each plot shows the one-to-one agreement.}
\label{fig:ew}
\end{figure}

We define $L(x)$ as the local flux for each line. $L(x) = F(x)$ in the range $x_{N}-25 \sigma_{N} < x < x_{N}+25 \sigma_{N}$ and zero elsewhere. We perform a bounded value  non-linear least square fit to the local flux $L(x)$ with a two component model $r(x)$, such that $r(x) = p(x) + g(x)$ where
\begin{equation}
\eqnum{3A}
p(x) = a + bx
\end{equation}
is a linear fit that models the underlying continuum and
\begin{equation}
\eqnum{4A}
g(x) = \displaystyle\sum_{i=1}^n a_{i} e^{\frac{(x-x_{oi})^2}{2 \sigma^2}}
\label{eq:gauss}
\end{equation}
is a gaussian that models the emission line, where $a_{L}, x_{oL}$ and $\sigma_{L}$ are defined as the gain, center and sigma of the gaussian fit to $L(x)$ respectively. We note that in the case of $H\beta$, due to possible underlying stellar absorption, we fit continuum between $4775-4815\AA$ and $4905 - 4945\AA$. It should be noted that $\sigma$ in equation~\ref{eq:gauss} does not have a summation index. If $n>1$ then the multiple gaussians simultaneously fitted will have different $a_{L}$ and $x_{oL}$ but the same $\sigma_{L}$. The parameters of the gaussian are constrained such that $a_{L}\geq0$, $x_{oL}$ can only vary by $\pm 2 \AA$ from the line centers obtained from NIST\footnote{This is done to account for possible errors in the redshift determination, reported line centers from NIST and/or wavelength calibrations} and $0<\sigma_{L}<10\AA$. For the [OII] $\lambda\lambda$3726, 3728 doublet  n=2 and for all other lines n=1. This enables us to fit the local continuum, $p(x)$, to all lines separately, except in the case of the [OII] doublet where we fit the continuum to both lines simultaneously.

We obtain a continuum-normalized local flux, $R(x)$, such that
\begin{equation}  
\eqnum{5A}
R(x) = \frac{L(x)}{p(x)} - 1.
\end{equation}
We refit a model, $g_{R}(x)$, of the form given in equation~\ref{eq:gauss} with n=3 for the $H_\beta$ and [OIII] $\lambda\lambda$4959, 5007 lines and n=2 for the [OII] $\lambda\lambda$3726, 3728 doublet. These values of n for the two sets of lines ensures that the fitted $\sigma$ will be the same for all lines in the set. We obtain a pseudo-EW for each line, $\mathrm{EW}_{p}$, such that $\mathrm{EW}_{p} = a_{R}\sigma_{R} \sqrt{2 \pi}$, where $a_{R}$ and $\sigma_{R}$ are parameters from the fit of $g_{R}(x)$ to $R(x)$. 

It becomes problematic to measure an equivalent width or pseudo-EW when the S/N is low in the underlying continuum. In this situation, the continuum is poorly measured and therefore a normalization to the continuum is prone to large errors.  Before normalizing the local flux, $L(x)$, to the continuum, $p(x)$, we estimate the noise of the local continuum, $\sigma_{c}$, by taking the rms of $C^\prime(x)$, where $C^\prime(x) = L(x) - r(x)$. We require that $p(x)$, in the region of the line $x_{o}-3\sigma_{L}<x<x_{o}+3\sigma_{L}$, be 1$\sigma_{c}$ above zero. If this condition is not met, we do not normalize to the continuum but instead fit a function of the form given in equation~\ref{eq:gauss} to $D(x)$, where $D(x) = L(x) - p(x)$. We then calculate a pseudo-EW from the parameters of the fit. In this study a small fraction ($\sim\!1\%$) of our galaxies have such low S/N in the continuum and are excluded from the study. 

\subsection{Comparison of line ratio determination}

\begin{deluxetable}{lcccc}

%% Keep a portrait orientation
\tablewidth{200pt}

%% Over-ride the default font size
%% Use Default (12pt)

%% Use \tablewidth{?pt} to over-ride the default table width.
%% If you are unhappy with the default look at the end of the
%% *.log file to see what the default was set at before adjusting
%% this value.

%% This is the title of the table.
\tablecaption{RMS deviation between pseudo-EW and flux calibrated line ratios}
%% This command over-rides LaTeX's natural table count
%% and replaces it with this number.  LaTeX will increment 
%% all other tables after this table based on this number
%\tablenum{1}

%% The \tablehead gives provides the column headers.  It
%% is currently set up so that the column labels are on the
%% top line and the units surrounded by ()s are in the 
%% bottom line.  You may add more header information by writing
%% another line between these lines. For each column that requries
%% extra information be sure to include a \colhead{text} command
%% and remember to end any extra lines with \\ and include the 
%% correct number of &s.
\tablehead{\colhead{} & \colhead{$[OII]/H_{\beta}$} & \colhead{$[OIII]/H_{\beta}$} & \colhead{$O_{32}$} & \colhead{$R_{23}$} \\ 
\colhead{} & \colhead{(dex)} & \colhead{(dex)} & \colhead{(dex)} & \colhead{(dex)} } 

%% All data must appear between the \startdata and \enddata commands
\startdata
KP03 & 0.11 & 0.05 & 0.12 & 0.08 \\
HJZ & 0.10 & 0.06 & 0.12 & 0.08 \\
%$\sigma$ & 0.06 & 0.02 & 0.06 & 0.04 \\
\enddata

%% Include any \tablenotetext{key}{text}, \tablerefs{ref list},
%% or \tablecomments{text} between the \enddata and 
%% \end{deluxetable} commands
\label{tab:comp}
%% General table comment marker

\tablecomments{RMS deviation comparison of one-to-one agreement between pseudo-EW and flux calibrated line ratios from \citet{Kobulnicky2003b}(top row) and our automated method (bottom row).}
%\tablecomments{RMS deviation comparison of one-to-one agreement from \citet{Kobulnicky2003b} (top row), our automated method (middle row) and the mean of the errors determined from the statistical measurement uncertainties of the spectra (bottom row).}

\end{deluxetable}

\begin{figure}
\begin{center}
\includegraphics[width = .7\columnwidth, keepaspectratio=1]{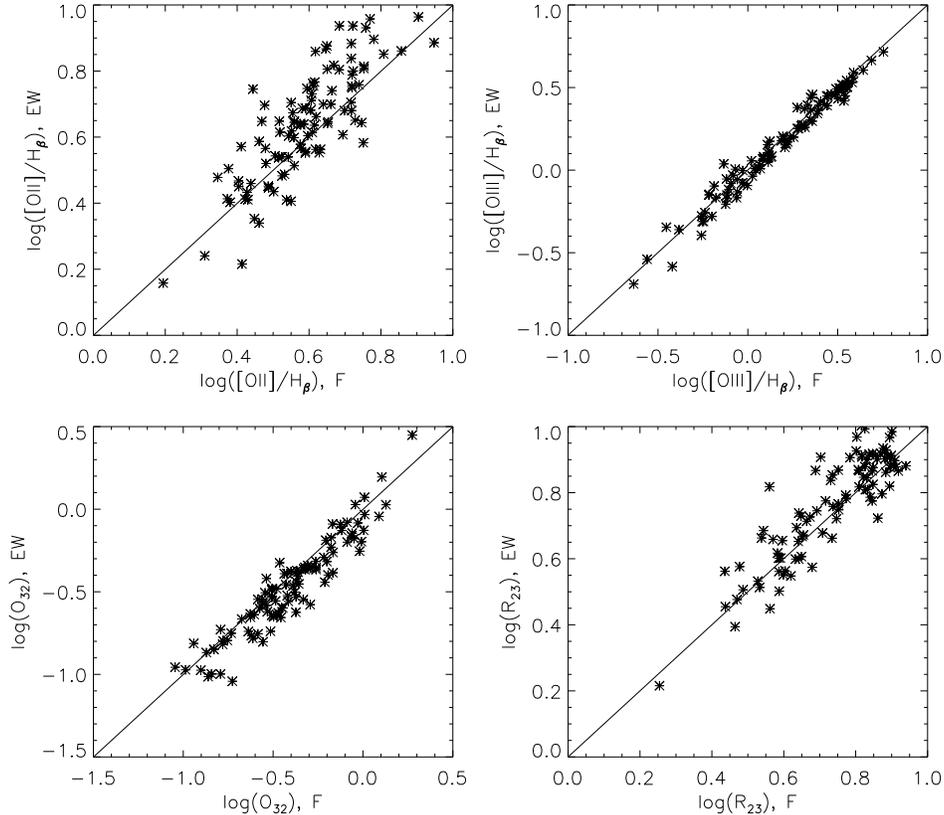}
\end{center}
\caption{Comparison of fitted emission line pseudo-EW ratios (denoted 'EW', y axis) to dereddened emission line flux ratios (denoted 'F', x axis) for 102 galaxies from the Nearby Field Galaxies Survey \citep{Jansen2000}. From top left panel going clockwise we show the $[OII]/H_{\beta}$, $[OIII]/H_{\beta}$, $R_{23}$ and $O_{32}$ ratios respectively. The solid line in each panel shows the one-to-one agreement between the two methods.}
\label{fig:nfgs}
\end{figure}

In order to test the robustness of our fitting method, we compare our pseudo-EW and line ratios with published equivalent widths and line ratios determined from calibrated fluxes from the Nearby Field Galaxies Survey (NFGS) \citep{Jansen2000}. KP03 compare how well interactively determined equivalent width ratios correspond to those determined from measured fluxes. We deredden the fluxes from NFGS sample as described in KP03 and use their results as a benchmark for comparison. Of the 198 galaxies in the survey, 118 have published fluxes. Of those 118, 104 galaxies have a S/N of at least 8 for [OII] $\lambda\lambda$ 3726, 3728, [OIII] $\lambda\lambda$ 4959, 5007 and $H_{\beta}$ emission lines. We remove two additional galaxies that have a good fit determined visually but poor agreement with ratios determined from calibrated fluxes, presumably due to the underlying continuum which is poorly measured, leaving us with a sample of 102 galaxies plotted in figures~\ref{fig:ew} and~\ref{fig:nfgs}.

Figure~\ref{fig:ew} shows a direct comparison between the equivalent widths determined by \citet{Jansen2000} and the pseudo-EW determined by our automated method. The rms dispersion from one-to-one agreement is 0.6, 4.0 and 0.7$\AA$ with the mean value of the equivalent width being 7.5, 28.0 and 15.3$\AA$ for $H_{\beta}$, [OII] and [OIII] respectively.

In order to justify the fact that the pseudo-EW can be used in inferring gas-phase metallicities, we need to establish agreement between the calibrated flux ratios and our pseudo-EW ratios. Figure~\ref{fig:nfgs} shows agreement in line ratios determined with our method as compared to those from calibrated fluxes with a rms dispersion from one-to-one agreement of 0.10, 0.06, 0.11 and 0.08 dex for the $[OII]/H_{\beta}, [OIII]/H_{\beta}, O_{32}$ and $R_{23}$ ratios respectively. Table~\ref{tab:comp} gives the rms deviation from one-to-one agreement between line ratios determined using the two different methods. The top row shows the rms deviation from KP03 and the bottom row shows the rms deviation for our automated method for our sample of 102 galaxies. The median absolute deviation between the equivalent width and flux calibrated metallicities is $<0.01$ dex.  In this local sample the errors from the statistical measurement uncertainties are about one half of those introduced by using equivalent widths or our pseudo-EW in line ratio determinations. Because galaxies are fainter at higher redshift, we expect them to have larger measurement uncertainties, thereby making the errors from using equivalent widths comparable or even less than the statistical measurement uncertainties.

\begin{figure*}
\includegraphics[scale=.65]{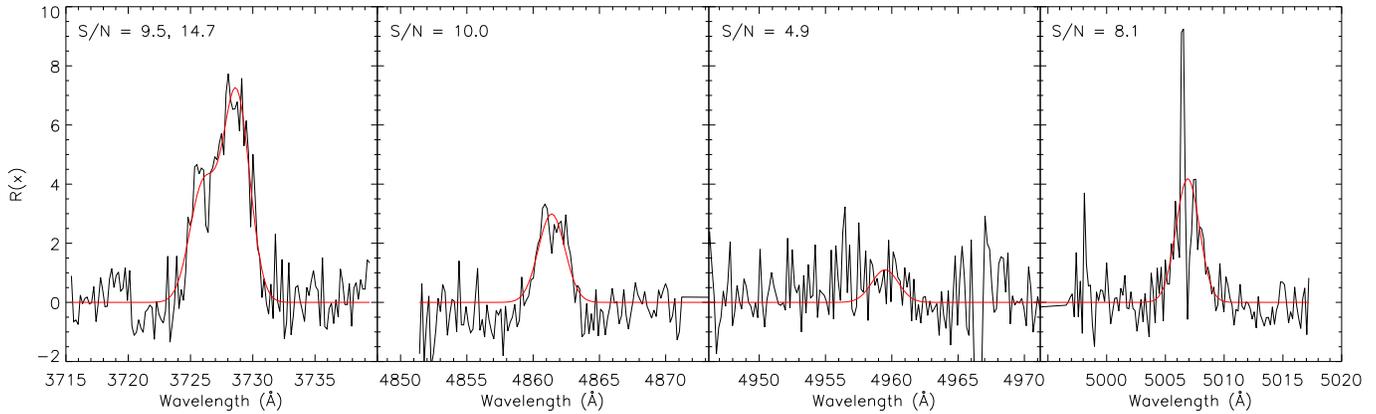}
\caption{A typical spectra from the DEEP2 survey with a  S/N $EW(H_{\beta})$ equal to our median value of 10.0 . From left to right, the spectra (black curve) and gaussian fit (red curve) for [OII] $\lambda\lambda$3726, 3728, $H_{\beta}$, [OIII]$\lambda$4959 and [OIII]$\lambda$5007. The S/N for each spectrum is displayed. The underlying Balmer absorption can be seen in the fit to $H\beta$.}
\label{fig:lsn}
\end{figure*}

The $O_{32}$ and $R_{23}$ ratios are of particular importance because of their use in inferring metallicities. Figure~\ref{fig:nfgs} and Table~\ref{tab:comp} demonstrate that our method for determining line ratios are comparable to those that use flux calibrated ratios. Figure \ref{fig:lsn} shows a typical spectrum and fit used in this study from the DEEP2 survey. Our automated method has several advantages over interactive determination of equivalent widths. 1) The method is automated so considerably faster than interactively determining equivalent widths and much more practical when working with large samples. 2) The automated method yields results that are reproducible because it no longer requires the input of wavelength limits in determining equivalent widths. 3) The measurement errors in flux, when available, can be propagated through to the determination of metallicity and therefore provide a much more robust estimate of the true errors in 12 + log[O/H]. 4) In data where a low S/N makes it difficult to distinguish between the line and the continuum, the automatic fit is less susceptible to confusion. 5) Low S/N spectra where the continuum is not detected, equivalent widths cannot be determined. These cases are handled in a much more robust manner allowing us to measure a pseudo-EW even when equivalent widths cannot be measured. 

\subsection{Redetermined Local MZ and LZ Relations from SDSS}

The local MZ relation has been determined from the SDSS data by both T04 and KE08. T04 derive a galaxy averaged gas-phase metallicity by simultaneously fitting the most prominent emission lines with stellar population synthesis and photoionization models. KE08 determine the metallicity using several common strong line diagnostics and calibrations and provide conversions between the various methods. Additionally, they investigate the effects of various AGN classification scheme and aperture covering fraction. The two determinations differ by as much as 0.3 dex when compared directly due mainly to the difference in methodology of determining metallicities. Converting to the same metallicity diagnostic using Table 3 from KE08 the difference decreases to $<0.05$ dex. 

The different MZ relations found in the literature at various redshifts may not necessarily be directly comparable. The differences resulting from the various metallicity diagnostics and calibrations have ostensibly been resolved by KE08. It should be noted however that it still remains uncertain whether relations calibrated using local galaxy samples are valid at higher redshifts. Care must be taken when comparing samples as selections effects lead to non-trivial differences in the determination of the MZ relation. We attempt to manage some of the differences arising from selection bias and different diagnostics by redetermining the local MZ relation from the SDSS data release 7 \citep[http://www.mpa-garching.mpg.de/SDSS/DR7/]{Abazajian2009} using similar selection cuts and the same metallicity calibration as our DEEP2 dataset.

The SDSS DR7 consists of $\sim 818,000$ galaxies spanning $0<z<0.7$. The emission line equivalent widths and galaxy stellar masses are calculated by the MPA/JHU group and are publicly available on the SDSS DR7 website. The equivalent widths have been corrected for stellar absorption. The $\emph{ugriz}$ fiber and C model magnitudes (recommended for galaxy photometry) are also provided. We determine the stellar masses by fitting the photometry to population synthesis models using the Le Phare code (see $\S\,3.1$). 

We select galaxies by constraining the redshift such that $0.04<z<0.1$. The lower limit is imposed to minimize aperture effects. Furthermore, we follow the method of KE08 whereby we determine the g-band fiber covering fraction by comparing the fiber and model magnitudes. KE08 determine that a $20\%$ covering fraction is insufficient to avoid aperture bias. We require a larger aperture fiber covering fraction of $>30\%$ in order to further reduce aperture effects. For the selected sample the median covering fraction is $38\%$. The upper redshift limit is imposed in order to minimize evolutionary effects and because the star-forming sample is found to be incomplete at higher redshifts \citep{Kewley2006}.

In order to minimize selection biases when comparing, we impose a similar selection criteria as applied to the DEEP2 sample. We require that SN $H\beta>3, \sigma_{r23} < 2$ and $EW(H\beta)>4\AA$. The SDSS data have much broader spectral coverage. For all galaxies in our redshift range we use the star-forming galaxies defined by the \citet{Kewley2006} classification scheme.   KE08 showed that MZ relation derived from the SDSS data using the $log([NII]/O[II])$ ratio is insensitive to the classification scheme used to identify and remove AGN. The selected SDSS data sample consists of $\sim21,000$ galaxies. The median $EW(H\beta)$ and S/N $H\beta$ of our selected sample is 8.9 and 20.7 respectively.

\begin{figure*}
\includegraphics[scale=0.78]{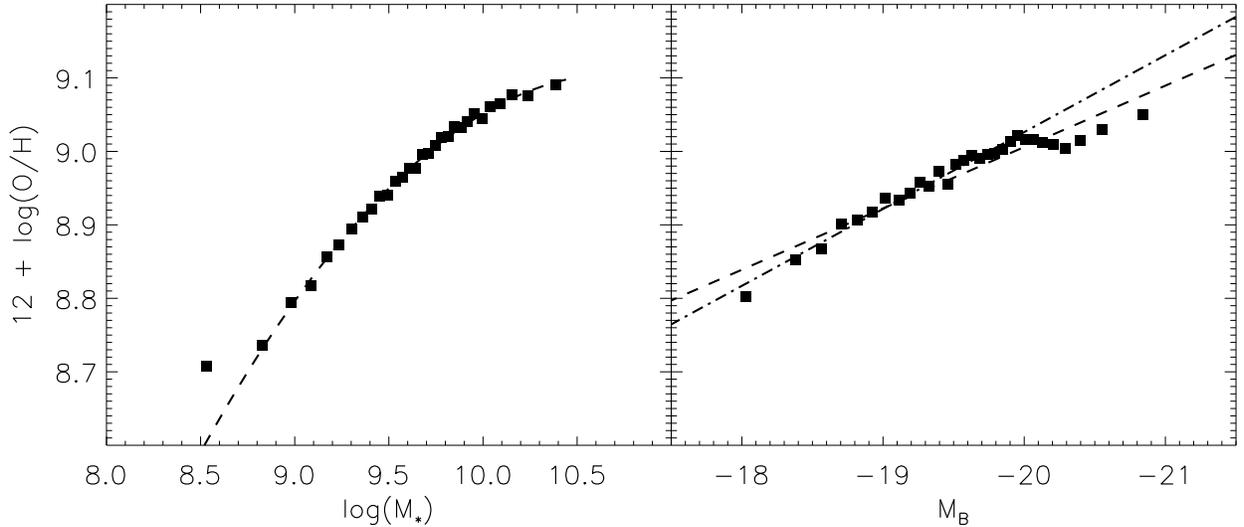}
\caption{The local MZ (left) and LZ (right) relation from $\sim21,000$ galaxies in the SDSS. The squares in each plot are the median metallicity for data sorted into 30 mass bins. The dashed line in the left panel is a quadratic fit to the binned data. The dashed line in the right panel is a linear fit to the binned data. The dot-dashed line in the right panel is a linear fit to the data with $M_{B}>-20$.}
\label{fig:mzsdss}
\end{figure*}

Figure \ref{fig:mzsdss} shows the local MZ and LZ relations derived from the selected SDSS sample. The metallicities have been derived using the equivalent widths as described in $\S \, 3.3$. We place galaxies on the lower $R_{23}$ branch when $log([NII]/O[II]) < -1.2$ (KE08). An insignificant fraction (25/21816) are on the lower branch. The fits to the relations use the same bootstrapping method described in section 4. For the MZ relation, the data are binned in 30 equally populated mass bins. From the left panel of Figure \ref{fig:mzsdss} it can be seen that the lowest mass bin suggests a turnover in the relation. This may be a result of incompleteness, though this is difficult to establish from these data. Whatever the case, the lowest mass bin is excluded when fitting the MZ relation. The MZ relation is given by
\begin{equation}
12 + log(O/H) = (9.051 \pm 0.001) + (0.151 \pm 0.004) \, x - (0.104 \pm 0.006) \, x^2,
\label{eq:fit}
\end{equation}
where $x = log(M_{\ast}) - 10$, the logarithm of the stellar mass in solar mass units zero pointed to reduce the covariance of the parameters by subtracting 10. The dashed line in the left panel of Figure \ref{fig:mzsdss} shows this fit.

The LZ relation is determined by sorting the data into 30 equally populated luminosity bins. The relation is shown in the right panel of Figure \ref{fig:mzsdss}. The dashed curve is a linear fit to the data and is parameterized by 
\begin{equation}
12 + log(O/H) =  (9.089\pm 0.003) - (0.0833 \pm 0.002)X_{B},
\label{eq:linlocal}
\end{equation}
where $X_B = M_B + 21$. The data turn over at $M_{B} \sim -20$, presumably due to the more fundamental turnover observed in the MZ relation at the higher masses. If we fit the LZ relation in the linear part of the relation, $M_{B} > -20$, the fit is
\begin{equation}
12 + log(O/H) =  (9.131\pm 0.006) - (0.105 \pm 0.003)X_{B}.
\label{eq:linlocal}
\end{equation}
This dot-dashed line in the right panel of Figure \ref{fig:mzsdss} shows this fit.

\bibliographystyle{apj}

\bibliography{metal.bib}

 \end{document}